\documentclass[submission, PhysicsCodebases]{SciPost}

\graphicspath{{figs/}}

\hypersetup{
    colorlinks,
    linkcolor={red!50!black},
    citecolor={blue!50!black},
    urlcolor={blue!80!black}
}

\usepackage[utf8]{inputenc}

\usepackage{physics}
\usepackage{braket}
\usepackage{orcidlink}
\usepackage{jlcode}
\usepackage{listings}
\usepackage{microtype}

\def\jlbasicfont{\ttfamily\footnotesize\selectfont}

\lstdefinestyle{code}{
  language=Julia,
  showstringspaces=false,
  keywordstyle={[1]\color{jlkeyword}\bfseries},
 keywordstyle={[2]\color{jlliteral}},
 keywordstyle={[3]\color{jlbase}},
 keywordstyle={[4]\color{jlmacros}},
  keywordstyle={[5]\color{jlbase}},
  commentstyle={\color{jlcomment}},
  stringstyle={\color{jlstrnum}},
  identifierstyle={\color{jlbase}},
  columns=fullflexible,
  keepspaces=true
}

\lstdefinestyle{codenohighlight}{
  language=Julia,
  showstringspaces=false,
  keywordstyle={[1]\color{black}},
  keywordstyle={[2]\color{black}},
  keywordstyle={[3]\color{black}},
  keywordstyle={[4]\color{black}},
  keywordstyle={[5]\color{black}},
  commentstyle={\color{black}},
  stringstyle={\color{black}},
  identifierstyle={\color{black}},
  columns=fullflexible,
  keepspaces=true
}

\lstnewenvironment{jlcodeblock}[1][]{
  \lstset{style=code,xleftmargin=0.2cm,xrightmargin=0.2cm,#1}
}{}
\lstnewenvironment{jlcodeblocknohighlight}[1][]{
  \lstset{style=codenohighlight,xleftmargin=0.2cm,xrightmargin=0.2cm,#1}
}{}

\newcommand{\codelink}[1]{\lstinputlisting[style=code]{#1}
  \noindent\begin{center}
  \filename@parse{#1}
  \href{\codeurl/\filename@base.\filename@ext}
  {\textcolor{blue}{\codelinktext}}
  \end{center}
}
\newcommand{\linkonly}[1]{
  \protect\filename@parse{#1}
  \href{\codeurl/\filename@base.\filename@ext}
  {\textcolor{blue}{\tt[\protect\filename@base.\protect\filename@ext]}}
}

\newcommand{\inline}{\jlinlcustomnohighlight}

\newcommand{\jlinlcustomnohighlight}[1]{%
    \settowidth{\jlinlem}{\jlinlfont{m}}%
    \lstinline[
        language=julia,
        style=codenohighlight,
        basicstyle=\jlinlfont,
        basewidth=\jlinlem
    ]^^a7#1^^a7}

\newcommand{\dt}{\delta\tau}

\newcommand{\Norm}[1]{\left\lVert #1 \right\rVert_1}
\newcommand{\E}{\mathbb{E}}

\newcommand{\ie}{\textit{i}.\textit{e}., }
\newcommand{\eg}{\textit{e}.\textit{g}.\ }

\DeclareSymbolFont{usualmathcal}{OMS}{cmsy}{m}{n}
\DeclareSymbolFontAlphabet{\mathcal}{usualmathcal}

\begin{document}

\begin{center}{\Large \textbf{
Rimu.jl: Random integrators for many-body quantum systems
}}\end{center}

\begin{center}
Matija \v{C}ufar\textsuperscript{1}\orcidlink{0000-0003-0734-2719},
C J Bradly\textsuperscript{1,2}\orcidlink{0000-0002-5413-777X},
Ray Yang\textsuperscript{1,3}\orcidlink{0000-0001-6628-5166},
Elke Pahl\textsuperscript{4}\orcidlink{0000-0002-6685-1655} and
Joachim Brand\textsuperscript{1$\star$}\orcidlink{0000-0001-7773-6292}
\end{center}

\begin{center}
{\bf 1}  Dodd-Walls Centre for Photonic and Quantum Technologies,
New Zealand Institute for Advanced Study, Massey University,
Private Bag 902104, North Shore, Auckland 0745, New Zealand
\\
{\bf 2} Department of Mathematics, Swinburne University of Technology, Hawthorn, VIC 3122, Australia
\\
{\bf 3} Department of Chemistry, Washington University in St.~Louis, St.~Louis, MO 63130, USA
\\
{\bf 4} MacDiarmid Institute for Advanced Materials and
Nanotechnology, Department of Physics, University of Auckland, Auckland 1010, New Zealand
\\
${}^\star$ {\small \sf J.Brand@massey.ac.nz}
\end{center}

\begin{center}
\today
\end{center}


\section*{Abstract}
{\bf
\href{https://github.com/RimuQMC/Rimu.jl}{Rimu.jl} is a Julia package for solving many-body quantum problems with exact diagonalisation and projector Monte Carlo.
The core of the package is a matrix-free implementation of Hamiltonians and other operators and compact representation of Fock states, which together allow for efficient methods suitable for high-performance computing.
Rimu.jl includes a pure Julia implementation of the full configuration interaction quantum Monte Carlo (FCIQMC) algorithm. FCIQMC is a variant of projector Monte Carlo, which solves the time-independent Schr\"odinger equation to obtain stochastic representations of the ground and low-lying excited eigenstates of a Hamiltonian.
Rimu.jl is extensible and makes use of interfaces for much of its core capabilities, including Hamiltonians and observables.
The package includes several well-known pre-defined model Hamiltonians and tools for analysing the Monte Carlo time series data.
We describe the main features and design principles of Rimu.jl, including the FCIQMC algorithm, with an emphasis on the unique solutions deviating from earlier FCIQMC implementations and the parallelisation strategy employed. Benchmarks are presented for both shared- and distributed-memory parallelisation with up to 1,000 nodes and 55,000 CPU cores.
}

\vspace{10pt}
\noindent\rule{\textwidth}{1pt}
\tableofcontents\thispagestyle{fancy}
\noindent\rule{\textwidth}{1pt}
\vspace{10pt}

\section{Introduction}\label{sec:intro}

Calculating the ground state of a quantum many-body problem is an important problem with applications in many areas of physics and chemistry. As the Hilbert space dimension of such problems grows exponentially with system size, solving them numerically in a deterministic way becomes intractable already at modest system sizes. However, through the use of stochasticity in Monte Carlo methods, both time and memory requirements of solving a quantum many-body problem can be drastically reduced. The family of methods that use stochastic techniques to solve quantum problems is known as quantum Monte Carlo (QMC).
A common feature of the widely varying QMC methods is that they all stochastically sample the solutions to the (deterministic) problem in such a way that taking an average of the recorded samples yields a good approximation of the true solution~\cite{nightingaleQuantumMonteCarlo1998,beccaContents2017}. Projector Monte Carlo is a class of QMC methods that uses the repeated application of a transition operator to an arbitrary initial state to project out all states but the ground state.
The transition operator is applied stochastically, and depending on the transition operator used, the basis, and the exact details of the stochastic application, several different variants of projector Monte Carlo can be defined~\cite{umrigarObservationsVariationalProjector2015}. Some notable examples include Green's function Monte Carlo~\cite{kalosMonteCarloCalculations1962,kalosStochasticWaveFunction1966}, diffusion Monte Carlo~\cite{reynoldsDiffusionQuantumMonte1990}, auxiliary field quantum Monte Carlo~\cite{zhangQuantumMonteCarlo2003}, path integral ground state~\cite{sarsaPathIntegralGround2000}, and full configuration interaction quantum Monte Carlo (FCIQMC)~\cite{boothFermionMonteCarlo2009}.

FCIQMC was first developed for use in quantum chemistry~\cite{boothFermionMonteCarlo2009} to solve problems related to the electronic structure of molecules~\cite{clelandStudyElectronAffinities2011,dadayFullConfigurationInteraction2012,holmesHeatBathConfigurationInteraction2016,Deustua2018,weserChemicalInsightsElectronic2021}, solid state applications~\cite{Booth2013}, and Hubbard models~\cite{Dobrautz2018}. Although the algorithm has mostly been used in these areas, it has been successfully applied to problems relating to the physics of ultra-cold quantum gases~\cite{taylorBoundExcitedStates2025,alhyderLatticeBosePolarons2025,rammelmullerMagneticImpurityOnedimensional2023,yangPolaronDepletonTransitionYrast2022} and, due to its simplicity could be applied to find the extremal eigenpairs of more general square matrices (or linear operators). This can be efficient when the matrix is sufficiently sparse and the non-zero matrix elements can be calculated quickly as required. A further property that helps efficiency is when the targeted eigenvector is highly non-uniform across Hilbert space with a strong concentration of its amplitude on a very small subspace, as is often the case in quantum systems of interest.

A major advantage of FCIQMC over other projector Monte Carlo algorithms is that it can treat systems with a sign problem by leveraging walker annihilation without resorting to approximations like the fixed-node approximation~\cite{Reynolds1982a,Needs2010}, assuming a sufficiently large amount of resources is employed \cite{boothFermionMonteCarlo2009,spencerSignProblemPopulation2012,Shepherd2014}. While this is not always feasible or efficient, it was recently demonstrated that walker annihilation can mitigate the sign problem without introducing a bias in a practically infinite Hilbert space, namely in the example of the Fröhlich polaron problem \cite{taylorBoundExcitedStates2025}.
Another feature of FCIQMC that sets it apart from diffusion Monte Carlo~\cite{Needs2010} is that it has no time step error, such that no extrapolation to small time steps is needed.

Several variants of FCIQMC exist: The original formulation~\cite{boothFermionMonteCarlo2009}, which is no longer in use, used integer coefficients, where each unit of the coefficient was treated as an independent walker. This was later extended to real (floating-point) coefficients with a semistochastic formulation~\cite{petruzieloSemistochasticProjectorMonte2012a} where a part of the Hilbert space is treated exactly with the rest sampled stochastically. Further expansions of the algorithm include: the initiator approximation~\cite{clelandCommunicationsSurvivalFittest2010,ghanemAdaptiveShiftMethod2020}, which trades the sign problem for a small bias; importance sampling~\cite{umrigarDiffusionMonteCarlo1993,ghanemPopulationControlBias2021}, which reduces the inherent population control bias present in FCIQMC; methods of improving convergence speed~\cite{bluntPreconditioningPerturbativeEstimators2019a,neufeldAcceleratingConvergenceFock2020a};
heat bath sampling~\cite{holmesHeatBathConfigurationInteraction2016}, which samples matrix elements with probabilities proportional to their value; and the fast randomised iteration approach~\cite{limFastRandomizedIteration2017,greeneWalkersStochasticQuantum2019a}, which optimises the stochastic compression of a state vector. In addition to computing ground state properties, variants of FCIQMC can also be used to compute excited states~\cite{bluntExcitedstateApproachFull2015a}, density matrices~\cite{bluntDensitymatrixQuantumMonte2014}, or even real-time dynamics~\cite{mccleanClockQuantumMonte2015a}. There are several previous implementations of FCIQMC available~\cite{gutherNECINElectronConfiguration2020,spencerHANDEQMCProjectOpenSource2019,sgreene8Sgreene8FRIES2025}, all focused on electronic structure problems.

In this paper, we present Rimu.jl, an extensible Julia~\cite{Julia-2017} framework for working with quantum many-body systems with a focus on FCIQMC and exact diagonalisation. Other QMC flavours and algorithms can be added later, and some work in this direction is already under development. The main motivation for developing a new implementation of FCIQMC was to go beyond electronic structure problems and allow full flexibility in choosing the quantum statistics of the underlying basis states.
The implementation as a function library based on the \href{https://github.com/SciML/CommonSolve.jl}{CommonSolve.jl} interface makes full use of Julia's multiple-dispatch paradigm and just-in-time compilation capabilities to generate efficient code that is suitable for various architectures spanning from laptop computers to high-performance computing facilities.

The purpose of this paper is to provide background information about the design principles behind Rimu.jl and the main algorithms employed. We also explain the context of representing a quantum many-body system on a computer in the manner that is relevant for using Rimu.jl. Rimu.jl continues to be developed as an open-source software project. Documenting the many (evolving) features of Rimu.jl is not in the scope of this paper but is left to Rimu.jl's \href{https://RimuQMC.github.io/Rimu.jl/stable/}{online documentation}.
In the main part of this paper, we present the variant of the FCIQMC algorithm employed by Rimu.jl, which dynamically determines the size of the deterministic subspace. Further, we describe the data structures used and show results of numerical benchmarks measuring the effects of the various parameters exposed by Rimu.jl, and its performance on thread-parallel and MPI~\cite{mpi41}-distributed computer systems is described.

We have designed the package to work with more general quantum many-body Hamiltonians beyond the electronic structure problem. Specifically, our own research interests are motivated by ultra-cold atom systems. To facilitate this, Rimu.jl is defined to work with arbitrary mixtures of (possibly non-number conserving) bosons and fermions with arbitrary spins. However, Rimu.jl also supports electronic structure Hamiltonians of molecules (using the FCIDUMP format~\cite{Knowles1989} as an interface to quantum chemistry packages), which is based on  spin-$1/2$ fermions. Extending the capabilities beyond the currently implemented Fock-state bases is also supported through the implemented type hierarchies and interfaces.

The benefits of implementing the package in Julia are many, but in our case the most valuable is the multiple dispatch feature, which allows us to tweak the internals of the implemented algorithms without having to modify the main code base. This speeds up development, and makes it easy to experiment with the algorithm and develop prototypes. Moreover, it provides a lot of flexibility for the user to add or tweak functionality as desired. Multiple dispatch also allows us to use type-level metaprogramming to have the Julia compiler generate specialised code tailored to the simulation parameters. Additionally, the input file for Rimu.jl is simply a Julia script, which makes working with Rimu.jl easy.

This paper is organised as follows: In Sec.~\ref{sec:example}, we start with a usage example that showcases some of the main features of Rimu.jl while introducing some underlying concepts in the context of concrete example calculation of the ground state properties of a Bose-Hubbard model.
The following sections go deeper into various aspects of the package.
In Sec.~\ref{sec:fciqmc}, we discuss the variant of the FCIQMC algorithm employed by Rimu.jl in detail, starting with a deterministic version, and then introducing stochasticity.
In Sec.~\ref{sec:Observables}, we discuss the various statistical techniques that allow us to extract useful information from an FCIQMC computation. This includes various ways of estimating the energy of a system or its observables.
We briefly present the two major ways exact diagonalisation can be performed with Rimu.jl in Sec.~\ref{sec:exact-diagonalization}.
In Sec.~\ref{sec:implementation}, we dive into the implementation details of the data structures used for representing Fock states, operators and vectors. We discuss how these implementation details allow for efficient parallel Monte Carlo and exact diagonalisation calculations.
In Sec.~\ref{sec:numerical-results}, we present numerical results showing how algorithm parameters affect an FCIQMC calculation, and present benchmarks of the performance of Rimu.jl with thread-parallel and MPI-distributed calculations.
Finally, in Sec.~\ref{sec:conclusion}, we conclude the paper and present possible future directions for Rimu.jl.

\section{Solving a quantum many-body problem with Rimu.jl: Bose Hubbard model example}
\label{sec:example}

We begin with a short example of an FCIQMC computation with Rimu.jl alongside introducing some basic concepts and linking to the relevant parts of this paper every time a new part of the software is introduced. We will use the Bose-Hubbard model, which is a well-studied model in quantum many-body physics that captures the important properties of the superfluid--Mott insulator transition \cite{FisherPRB1989} and is relevant to ultra-cold atom experiments \cite{Bloch2012}. As a simple case, consider a one-dimensional lattice with periodic boundaries, realising a ring-shaped chain of $M$ lattice sites. In the language of second quantisation \cite{Sakurai2010a}, which is convenient for these models and used heavily in the Rimu.jl documentation, the Hamiltonian reads
\begin{equation}\label{eq:bose-hubbard-hamiltonian}
\hat{H} = -t\sum_{\langle i,j\rangle}\hat{a}^\dagger_i\hat{a}_j + \frac{u}{2}\sum_{i=1}^{M}\hat{n}_i(\hat{n}_i-1)\,,
\end{equation}
where $\hat{a}^\dagger_i$ ($\hat{a}_i$) creates (destroys) a bosonic particle on lattice site $i\in \{1,2,\ldots M\}$. The first term describes hopping of a single particle (with hopping strength $t$) to a neighbouring site, as
$\langle i,j\rangle$ denotes all pairs of adjacent sites.
The second term describes a pair-wise on-site interaction between two particles with strength $u$, where $\hat{n}_i = \hat{a}^\dagger_i\hat{a}_i$ represents the particle number on site $i$.

By choosing a suitable basis (here Fock states), we can convert objects that live in an abstract Hilbert space into the space of numbers [here $\mathbb{R}^D$, where $D=\binom{N+M-1}{N}$ and $N$ is the number of particles]. Determining the ground state then, conceptually, becomes finding the smallest eigenvalue and associated eigenvector of a matrix.
The Bose-Hubbard Hamiltonian of Eq.~\eqref{eq:bose-hubbard-hamiltonian} defines a matrix with the elements $\langle\mathbf{n}|\hat{H}|\mathbf{m}\rangle$ in a basis of Fock states
\begin{align} \label{eq:FockState}
    |\mathbf{n}\rangle \equiv |n_1,n_2, \ldots n_M\rangle
    \equiv \prod_{j=1}^M \frac{1}{\sqrt{n_j!}} \left(a_j^\dagger\right)^{n_j} |\mathrm{vac}\rangle .
\end{align}
Here $N = \sum_{j=1}^M n_j$ is the total particle number of the bosons, which is conserved by the Hamiltonian and will thus be regarded as fixed. It becomes a parameter of the calculation. Fock states of bosons with fixed particle number are represented by the type \inline{BoseFS\{N,M\}} in the code, which can be generated programmatically, or entered with two different notations
\begin{jlcodeblock}
BoseFS(1,3,0,5) == fs"|1 3 0 5⟩" # true
\end{jlcodeblock}
Fock states are important in Rimu.jl because they are used as addresses to compute the matrix elements of a Hamiltonian, and also for storing state-vector (wave function) data. The implementation of Fock addresses, which could also be of fermionic or composite nature, is covered in Sec.~\ref{ssec:addresses}.

In this example, we will look at a system of $N=12$ particles in $M=12$ sites, which results in a Hilbert space dimension of 1,352,078. This is small enough to be quickly solvable with Rimu.jl's FCIQMC and sparse eigenvalue solvers, but too large for dense numerical linear algebra. In the example, we will start with a ground state calculation with projector Monte Carlo, then extend it to demonstrate computing excited states and observables.
Along the way, we will compare the Monte Carlo results to those obtained with exact diagonalisation.

\subsection{Computing the ground state energy with projector Monte Carlo}\label{sec:example-basic}

If Rimu.jl is not yet installed, we install it through the Julia package manager by typing
\begin{jlcodeblock}
]add Rimu
\end{jlcodeblock}
in the Julia REPL.
We can now set up the computation by loading Rimu.jl, then creating a bosonic Fock address and defining the Hamiltonian:
\begin{jlcodeblock}
using Rimu
N = 12 # number of particles
M = 12 # number of sites
address = near_uniform(BoseFS{N,M}) # fs"|1 1 1 1 1 1 1 1 1 1 1 1⟩"
hamiltonian = HubbardRealSpace(address; u=6.0, t=1.0)
\end{jlcodeblock}
The Fock address is created first to define the quantum statistics and the size of the Hilbert space of the problem. The bosonic Fock state constructed with \inline{near_uniform} is one where the $N=12$ particles are uniformly distributed across the $M=12$ modes, which represent the lattice sites of the Hubbard model.
We then pass the address and the interaction parameter $u=6$ and hopping strength $t=1$ to the Hamiltonian constructor. This corresponds to strong on-site interactions typical of the Mott-insulating regime \cite{FisherPRB1989}. Passing the address to the Hamiltonian in this way sets the number of particles and lattice sites, as well as providing a starting point for the FCIQMC calculation. Section \ref{ssec:hamiltonians} discusses how Hamiltonians and operators work in Rimu.jl.

A projector Monte Carlo calculation proceeds by iteratively evolving a state vector in steps (or Monte Carlo steps, or imaginary time). Conceptually, a state vector is defined in a basis of Fock states $|\mathbf{n}\rangle$
\begin{align} \label{eq:statevector}
    |\Psi\rangle = \sum_\mathbf{n} c_\mathbf{n} |\mathbf{n}\rangle ,
\end{align}
where the coefficient vector $\{c_\mathbf{n}\}$ is sometimes also called the wave function. State vectors are represented in Rimu.jl using hash tables (dictionaries) that store both the Fock addresses and the coefficients, as described in Sec.~\ref{ssec:dvecs}. While the state vectors can be created and handled internally, they are available to be manipulated by the user.

The main quantity of interest is the ground state energy. In projector Monte Carlo it can be obtained (estimated) from different energy estimators, as discussed in Sec.~\ref{sec:EnergyEstimators}. Here we will collect data for the projected energy [defined by Eq.~\eqref{eq:ProjectedEnergy}]. It works by projecting the current state vector of the running Monte Carlo simulation to a reference vector.
\begin{jlcodeblock}
reference_vector = PDVec(address => 1.0)
post_step_strategy = ProjectedEnergy(hamiltonian, reference_vector)
\end{jlcodeblock}
We set up the reference vector  as a \inline{PDVec} (short for parallel dictionary vector, see Sec.~\ref{ssec:dvecs}) with the value of 1 at the starting address.

Now, we create the \inline{ProjectorMonteCarloProblem}, which holds the information necessary to start an FCIQMC run and solve it with the \inline{solve} function following the \href{https://github.com/SciML/CommonSolve.jl}{CommonSolve.jl} interface. We finally store the result to a \inline{DataFrame} from the \href{https://github.com/JuliaData/DataFrames.jl}{DataFrames.jl}~\cite{bouchet-valatDataFramesjlFlexibleFast2023} package and display it.
\begin{jlcodeblock}
problem_fciqmc_basic = ProjectorMonteCarloProblem(
    hamiltonian;
    last_step=10_000,
    target_walkers=100_000,
    post_step_strategy,
)

result = solve(problem_fciqmc_basic)
df = DataFrame(result)

display(df)
# 10000×11 DataFrame
#    Row │ step   len    shift     norm           exact_steps  inexact_st ⋯
#        │ Int64  Int64  Float64   Float64        Int64        Int64      ⋯
# ───────┼─────────────────────────────────────────────────────────────────
#      1 │     1      6  -1.83494     15.0                  0             ⋯
#      2 │     2      9  -2.42858     19.1719               0
#      3 │     3     12  -2.68342     23.3947               0
#      4 │     4     19  -3.51728     30.5274               0
#    ⋮   │   ⋮      ⋮       ⋮            ⋮             ⋮             ⋮    ⋱
#   9997 │  9997  48648  -7.79238      1.00169e5          329          48 ⋯
#   9998 │  9998  48399  -7.77228  99918.7                332          48
#   9999 │  9999  48347  -7.76695  99855.1                331          48
#  10000 │ 10000  48231  -7.75921  99763.2                331          48
#                                           6 columns and 9992 rows omitted
\end{jlcodeblock}
Here, we performed $10{,}000$ steps of the FCIQMC algorithm. The target walker number, which determines the desired one-norm of the state vector, was set to $100{,}000$. See Sec.~\ref{sec:fciqmc} for more details on the FCIQMC algorithm.

\begin{figure}[t!]
\includegraphics[width=\textwidth]{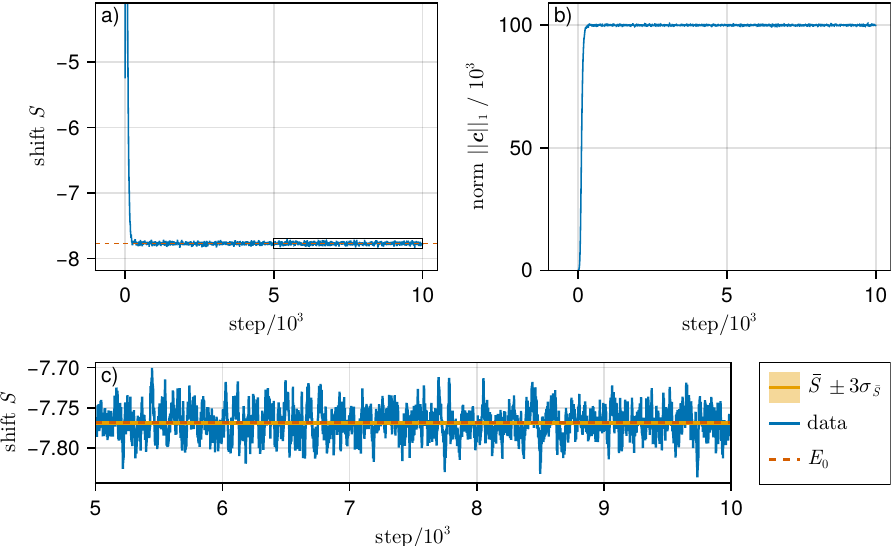}
\caption{\label{fig:timeseries2}
    FCIQMC time series data obtained in the example. Panel (a) shows the shift and panel (b) the one-norm, also referred to as the walker number. A close-up of the equilibrated region of the shift time series is shown in panel (c), which also shows the exact energy $E_0$ (dashed red line) and the estimate given by the mean of the shift $\bar{S}$ (orange line) with a blocked standard error $\sigma_{\bar{S}}$ (orange band).}
\end{figure}

The \inline{DataFrame} holds time series of the various statistics that Rimu.jl keeps track of. Of particular interest are the shift $S$ and the one-norm (labelled ``norm''), which are plotted in Fig.~\ref{fig:timeseries2}. The shift is used as a mechanism to control the norm of the FCIQMC state vector, as explained in detail in  Sec.~\ref{ssec:fciqmc:basic}. The long-time mean of the shift also acts as an energy estimator. Initially, the shift varies wildly, but then quickly settles to fluctuating around a fixed value. On the same time scale the one-norm of the FCIQMC vector grows and settles around the desired value of 100,000 walkers as is shown in panel (b) of Fig.~\ref{fig:timeseries2}. The value of the shift fluctuates around an estimator for the ground state energy of the Hamiltonian. For the estimate we discard the initial transient part of the time series and perform a blocked averaging  (see Sec.~\ref{sec:Observables} and Refs.~\cite{flyvbjergErrorEstimatesAverages1989b,jonssonStandardErrorEstimation2018}) to extract the mean $\bar{S}$ and its estimated statistical error $\sigma_{\bar{S}}$. These quantities are shown in Fig.~\ref{fig:timeseries2} (c). There, we see that the true energy $E_0$ lies within the error bars given by the blocked average $\bar{S}$.

In the code, we perform this averaging through the \inline{shift_estimator} function.
\begin{jlcodeblock}
se = shift_estimator(df; skip=5000)
\end{jlcodeblock}
Here, \inline{skip} sets the number of steps to be discarded from the beginning of the time series.
Next, we extract the projected energy estimate by utilising the \inline{projected_energy} function
\begin{jlcodeblock}
pe = projected_energy(df; skip=5000)
\end{jlcodeblock}
The values given by these functions are $\bar{S} = -7.7686 \pm 0.0008$ and $\overline{E}_y = -7.7681 \pm 0.0004$ for the shift and projected energy, respectively. These values compare favourably to the exact ground state energy $E_0 \approx -7.76815$ obtained by exact diagonalisation (as discussed later, see Sec.~\ref{sec:example-exact}).
Section \ref{sec:Observables} covers the calculation of observables including a discussion of the blocked averaging procedure used for decorrelating time series data.

\subsection{Excited states}\label{sec:example-excited}

Beyond ground states, Rimu.jl also supports computing consecutive low-lying excited states, which are obtained by Gram-Schmidt orthogonalisation \cite{bluntExcitedstateApproachFull2015a}.
\begin{jlcodeblock}
problem_fciqmc_excited = ProjectorMonteCarloProblem(
    hamiltonian;
    last_step=10_000,
    target_walkers=100_000,
    post_step_strategy,
    n_spectral=2,
)

result = solve(problem_fciqmc_excited)
df = DataFrame(result)
\end{jlcodeblock}
Setting \inline{n_spectral=2} triggers two independent step-synchronised FCIQMC simulations where the second's instantaneous state vector is orthogonalised to the first one in order to target the first excited state.
The resulting \inline{DataFrame} now has a larger number of columns. For most of the statistics calculated, it contains columns postfixed with \inline{_r1s1} and \inline{_r1s2}, which correspond to the first and second spectral states.
To get an estimate of the excited state energy, we again use the \inline{shift_estimator} function, but now have to give it the name of the shift column from the \inline{DataFrame}.
\begin{jlcodeblock}
se_state_2 = shift_estimator(df; shift=:shift_r1s2, skip=5000)
\end{jlcodeblock}
The resulting value $\bar{S_2} = -5.665 \pm 0.001$ matches the true excited state energy value of $E_1 \approx -5.6671$ well.
While the ground-state calculations for the Bose-Hubbard model are free of the Monte Carlo sign problem \cite{spencerSignProblemPopulation2012}, this is not the case any more for excited states, which always contain nodes due to orthogonality to the ground state.
Computing excited state energies can thus be challenging in practice even in systems that are free of the Monte Carlo sign problem for ground state calculations. Excited state calculations with Rimu.jl for the Fr\"ohlich polaron model were reported in Ref.~\cite{taylorBoundExcitedStates2025}, where the sign problem could be mitigated by walker annihilation.

\subsection{Observables}\label{sec:example-observables}

\begin{figure}[t!]
\includegraphics[width=\textwidth]{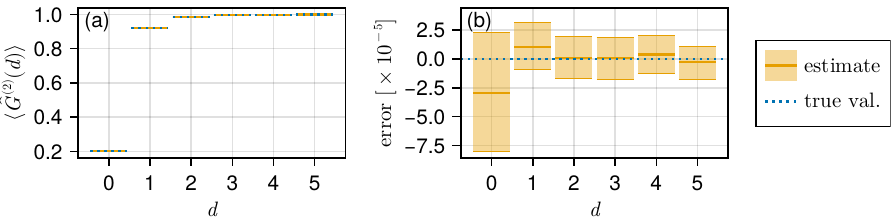}
\caption{\label{fig:g2-plot}Computing observables with FCIQMC. Estimates of the density-density correlations $\langle \hat{G}^{(2)}(d)\rangle$ over a range of distances of $d$ lattice sites. In panel (a), the orange bars show the estimates obtained from FCIQMC, where their widths are equal to $3\sigma$ error bars. The exact values are plotted as dotted red lines. Panel (b) shows the error calculated from the differences between the estimates and the exact values.}
\end{figure}

Observables that are defined by the expectation value of a general operator (different from the Hamiltonian) can be calculated with Rimu.jl as well. This typically requires the step-synchronised simulation of at least two statistically independent state vectors called \emph{replicas}. Here we demonstrate computing
density-density correlations as the expectation value of the operator
\begin{equation}\label{eq:g2}
  \hat{G}^{(2)}(d) = \frac{1}{M}\sum_{i}\hat{n}_{i}\left(\hat{n}_{i+d} - \delta_{d,0}\right)\,,
\end{equation}
which represents the density-density correlation at a distance of $d$ lattice sites.
Both Hamiltonians and observables are linear operators in Hilbert space and behave in a similar manner, which is reflected in Rimu.jl's type hierarchy as explained in Sec.~\ref{ssec:hamiltonians}. We set up the computation as follows:
\begin{jlcodeblock}
g2_operators = [G2RealCorrelator(d) for d in 0:5]
n_replicas = 2
replica_strategy = AllOverlaps(n_replicas; operator=g2_operators)

problem_fciqmc_observables = ProjectorMonteCarloProblem(
    hamiltonian;
    last_step=10_000,
    target_walkers=100_000,
    post_step_strategy,
    replica_strategy,
)

result = solve(problem_fciqmc_observables)

df = DataFrame(result)
\end{jlcodeblock}
Here, we first initialise a vector of observables with values of $d=0,\dots,5$ and then pass them into the \inline{replica_strategy} \inline{AllOverlaps}. The latter is a mechanism for computing overlaps between the state vector replicas and observables after every Monte Carlo step. In order to compute those without bias, at least two independent runs of FCIQMC (replicas) must be run concurrently. The number of replicas can be specified, but two replicas are run by default when using \inline{AllOverlaps}. See Sec.~\ref{sec:GeneralObservables} for more information on calculating observables with replicas. The resulting \inline{DataFrame} now contains columns containing the dot-products between the state vectors and each observable, which can be averaged with the \inline{rayleigh_replica_estimator} function.
\begin{jlcodeblock}
g2_d_is_0 = rayleigh_replica_estimator(result; op_name="Op1", skip=5000)
\end{jlcodeblock}
The results are shown in Fig.~\ref{fig:g2-plot} and compared with the exact values. As the model was set with unit filling (i.e.~the same number of particles as lattice sites) and sufficiently large interaction parameter $u/t=6$, the ground state realises a Mott insulating state (strictly speaking a precursor because of the finite system size). We find that the on-site correlation $\langle \hat{G}^{(2)}(0)\rangle$, i.e.~the probability of finding a second particle on the same site where a first particle is found is low, here around $20\%$. For the non-local density-density correlations the values are much higher and close to 1, which is exactly as expected in a Mott-insulating state.
Computing observables with Rimu.jl's FCIQMC was used to compute density-density correlations of an impurity in a one-dimensional interacting Bose gas reported in Ref.~\cite{yangPolaronDepletonTransitionYrast2022}.

\subsection{Exact diagonalisation}\label{sec:example-exact}

Because our example is small enough, we have been able to confirm the correctness of the FCIQMC result by performing an exact diagonalisation of the Hamiltonian and comparing the exact and numerical results for the energy. Rimu.jl's exact diagonalisation capabilities are discussed in Sec.~\ref{sec:exact-diagonalization}. The package is compatible with eigenvalue solvers from several external packages and implements suitable package extensions; here, we use the \href{https://github.com/Jutho/KrylovKit.jl}{KrylovKit.jl}~\cite{haegemanKrylovKit2025} package.
The interface for solving the Hamiltonian in Eq.~\eqref{eq:bose-hubbard-hamiltonian} with exact diagonalisation is similar to the one used for FCIQMC --- we first define a problem by instantiating an \inline{ExactDiagonalizationProblem} which we then solve with the \inline{solve} function.
\begin{jlcodeblock}
using KrylovKit

problem = ExactDiagonalizationProblem(
    hamiltonian; algorithm=KrylovKitSolver(false),
)
exact_result = solve(problem; howmany=2)
\end{jlcodeblock}
The exact energies $E_0$ and $E_1$ can then be extracted by accessing
\begin{jlcodeblock}
exact_result.values
# 2-element Vector{Float64}:
#  -7.76815...
#  -5.66708...
\end{jlcodeblock}
while the eigenvectors are stored in \inline{exact_result.vectors}:
\begin{jlcodeblock}
exact_result.vectors[1]
# 1352078-element PDVec: style = IsDeterministic{Float64}()
#   fs"|2 0 2 3 0 1 0 0 1 1 0 2⟩" => 3.72353e-6
#   fs"|0 1 0 0 0 0 2 2 1 2 3 1⟩" => 2.16686e-8
#   ⋮   => ⋮
\end{jlcodeblock}
Exact diagonalisation of transcorrelated~\cite{jeszenszkiEliminatingWavefunctionSingularity2020a} Hamiltonians with Rimu.jl was used to solve one-dimensional few-fermion and impurity problems in Refs.~\cite{backertEffectiveTheoryStrongly2025,rammelmullerMagneticImpurityOnedimensional2023}.

\subsection{Larger Hilbert space dimensions and the sign problem}
This example shows the basics of Rimu.jl's usage in a small system. Rimu.jl can be used in a similar manner with many other types of Hamiltonians and both its FCIQMC and exact diagonalisation capabilities can be scaled up to much larger problems. As an example, Rimu.jl was used in Ref.~\cite{alhyderLatticeBosePolarons2025} to compute the (sign-problem-free) ground state of a two-dimensional impurity-Bose-Hubbard Hamiltonian with 100 particles on 100 sites, which has a Hilbert space dimension of $\sim4.5\times10^{60}$.
In cases where a sign problem occurs, FCIQMC can sometimes overcome this by walker annihilation if the walker number is sufficiently large \cite{boothFermionMonteCarlo2009,spencerSignProblemPopulation2012}. This was achieved with Rimu.jl for excited states of the Fr\"ohlich polaron model in Ref.~\cite{taylorBoundExcitedStates2025}. The sign-problem suppressing initiator approximation \cite{clelandCommunicationsSurvivalFittest2010} is also implemented in Rimu.jl and was used for calculations in a momentum-space Bose-Hubbard model in Ref.~\cite{yangPolaronDepletonTransitionYrast2022}.
As the computational effort grows with system size and complexity of the problem, so does the demand for computational resources. Rimu.jl has extensive capabilities for leveraging both multi-core parallelism and multi-node distributed computing. The core implementation strategies are discussed in Sec.~\ref{ssec:parallel} and scaling benchmarks on a high-performance computing facility presented in Sec.~\ref{sec:numerical-results}.

\section{The FCIQMC algorithm}\label{sec:fciqmc}

FCIQMC is an algorithm for stochastically sampling the lowest eigenpair $(E_0, \vb{v}_0)$ of a linear operator $\hat{H}$.
Since we are dealing with finite systems, $\hat{H}$ can be represented as a matrix but, as we will discuss later, there is no need to store the actual matrix in memory. How operators and vectors are implemented in the code is discussed in more detail in Secs.~\ref{ssec:hamiltonians} and~\ref{ssec:dvecs}.

\subsection{A fixed point iteration}\label{ssec:fciqmc:basic}

FCIQMC belongs to the wider class of projector Monte Carlo methods. These methods work by discretising the imaginary-time Schr\"{o}dinger equation
\begin{equation}
  \frac{\partial\ket{\Psi(\tau)}}{\partial\tau} = -\hat{H}\ket{\Psi(\tau)}\,,
\end{equation}
and then solving the discretised equation stochastically.
In the case of FCIQMC, the discretisation employed is an Euler method
\begin{equation}\label{eq:discretised}
  \ket{\Psi(\tau + \dt)} = \ket{\Psi(\tau)} -\dt\hat{H}\ket{\Psi(\tau)}\,,
\end{equation}
where $\dt$ is a (small) imaginary time step.

Since we are solving the equation numerically, we realise $\hat{H}$ as a matrix $\vb{H}$ with entries
\begin{equation}
    H_{i,j} = \braket{i | \hat{H} | j}
\end{equation}
where $i,j$ are (not necessarily integer) indices and $\ket{i},\ket{j}$ are basis states. Similarly, we realise the state after $n$ steps $\ket{\Psi(n\,\dt)}$ as a coefficient vector $\vb{c}^{(n)}$ with coefficients
\begin{equation}
    c_i^{(n)} = \braket{i | \Psi(n\,\dt)}\,.
\end{equation}
We use $\vb{c}^{(0)}$ to denote the starting vector.
Then, Eq~\eqref{eq:discretised} can be converted to
\begin{equation}\label{eq:fciqmc}
  \vb{c}^{(n + 1)} = \vb{T}^{(n)} \vb{c}^{(n)}\,,
\end{equation}
where $\vb{T}$ is the FCIQMC transition operator
\begin{equation}\label{eq:propagator}
  \vb{T}^{(n)} := \vb{1} + \dt\ \left(S^{(n)}\vb{1} - \vb{H}\right)\,.
\end{equation}
Here, $\vb{1}$ is the identity matrix and $S^{(n)}$ is an energy \textit{shift}. 
If the shift is held constant, the scheme admits exponential solutions, where the norm of $\vb{c}$ grows or decays exponentially, with  $S^{(n)}=E_0$ being the only stable solution. Hence, Eq.~\eqref{eq:fciqmc} presents an unstable fixed-point iteration scheme on $\vb{c}$ with a fixed point at the desired eigenpair $\vb{c} =\vb{v}_0$ and $S = E_0$.

To stabilise the fixed point, we update the value of $S^{(n)}$ after every step with the following procedure~\cite{yangImprovedWalkerPopulation2020}
\begin{equation}\label{eq:shift-update}
  S^{(n+1)} = S^{(n)} -\frac{\zeta}{\dt}\ln
      \frac{\Norm{\vb{c}^{(n+1)}}}{\Norm{\vb{c}^{(n)}}}
  - \frac{\xi}{\dt}\ln\frac{\Norm{\vb{c}^{(n+1)}}}{N_{\mathrm{t}}}\,,
\end{equation}
where $N_\mathrm{t}$ is the target 1-norm of the coefficient vector and $\zeta$ and $\xi$ are parameters controlling the dynamics of the shift.
It was shown in Ref.~\cite{yangImprovedWalkerPopulation2020} that the combination of Eq.~\eqref{eq:fciqmc} with the shift update of Eq.~\eqref{eq:shift-update} behaves like a (discretised) damped harmonic oscillator differential equation for the logarithmic 1-norm of the vector. In this interpretation of the equation, the terms with parameters $\zeta$ and $\xi$ represent the damping and forcing terms, respectively. The default choice of $\xi=\zeta^2/2$ corresponds to a critically damped harmonic oscillator, and provides a time scale of $\tau_c = (2/\zeta)\delta\tau$ on which fluctuations of the norm are damped towards the target norm $N_\mathrm{t}$. The combination of Eqs.~\eqref{eq:fciqmc} and~\eqref{eq:shift-update} presents a stable fixed-point iteration scheme on the pair $(S^{(n)}, \vb{c}^{(n)})$ with a fixed point at the eigenpair $(E_0, \vb{v}_0)$ with a normalized eigenvector $\Norm{\vb{v}_0} = N_{\mathrm{t}}$.

An advantage of FCIQMC over some other methods is that it has no time step error. This can be seen from the fact that $\dt$ does not play a role in determining the values of the fixed points. However, the choice of $\dt$ is still important, as the algorithm will only converge if
\begin{equation}
  \dt < \frac{2}{E_{\max} - E_0},
\end{equation}
where $E_{\max}$ is the largest eigenvalue of $\vb{H}$ and $E_{\max} - E_0$ is its spectral range~\cite{spencerSignProblemPopulation2012}.

As demonstrated in Sec.~\ref{sec:example-excited}, FCIQMC can be extended to compute low-lying excited states in addition to the ground state. To compute the first $K$ states, we run FCIQMC in parallel on $K$ independent vectors $\vb{c}^{(n,k)}, k=0,\dots,K-1$ with independent shifts $S^{(n,k)}$. Then after every step, we orthogonalise the vectors using the Gram-Schmidt procedure, which ensures each vector converges to a separate spectral state $\vb{v}_k$, while each shift converges to the energy of that state $E_k$~\cite{bluntExcitedstateApproachFull2015a}.

\subsection{Quantum Monte Carlo}\label{ssec:fciqmc:montecarlo}

The method described thus far is equivalent to the deterministic power method and as such provides no benefits over standard methods for eigenvalue computation. However, unlike many other methods, it can be performed stochastically by replacing the operator $\vb{T}$ with a stochastic operator $\check{\vb{T}}$. For this to be beneficial, we pick $\check{\vb{T}}$ in a way that reduces the computational cost of the operator application in Eq.~\eqref{eq:fciqmc}. At the same time, we must ensure that the fixed points of Eqs.~\eqref{eq:fciqmc} and~\eqref{eq:shift-update} remain unchanged. To achieve this, we require the stochastic operator to be equal to the deterministic operator in expectation:
\begin{equation}
\E\left[\check{\vb{T}}^{(n)}\vb{c}^{(n)}\right] = \vb{T}^{(n)}\vb{c}^{(n)}\,,
\end{equation}
where $\E$ denotes the ensemble average.

Because applying $\check{\vb{T}}$ is noisy, the algorithm does not cleanly converge to $(E_0, \vb{v}_0)$. Instead, the output is a time series with two phases; an equilibration phase, where the algorithm moves $S^{(n)}$ and $\vb{c}^{(n)}$ towards the desired fixed points, and a phase where the algorithm has reached the fixed point and is fluctuating around it. To extract an estimate of $E_0$ from the algorithm, we discard all shift data from the first phase and take an average of the second. An example of an FCIQMC time series is shown in Fig.~\ref{fig:timeseries2}.

Rimu.jl provides several ways of realising the stochastic operator $\check{\vb{T}}$. In this section, we describe the one we have found to be the most useful.
Although the algorithm was originally formulated in Monte Carlo terminology with walkers who spawn, die, and annihilate~\cite{boothFermionMonteCarlo2009}, here we describe the procedures in terms of stochastic \textit{matrix compression} and \textit{vector compression}, as introduced in Ref.~\cite{limFastRandomizedIteration2017}.

To start, we write a general coefficient vector $\vb{c}$ as
\begin{equation}
  \vb{c} \equiv \sum_i c_i \vu{i}\,.
\end{equation}
Here, we can call the orthonormal set of vectors $\{\vu{i}\}_i$ the primitive basis indexed by the addresses $i$. In Rimu.jl, these addresses represent Fock states, Eq.~\eqref{eq:FockState}. A matrix-vector product can then be seen as a weighted sum of the matrix columns
\begin{equation}\label{eq:spawning}
   \vb{T}\,\vb{c} \equiv \sum_i\vb{T}_{:,i}\,c_i\ \equiv \sum_{j,i} T_{j,i} \, c_i\,\vu{j}\,,
\end{equation}
where $\vb{T}_{:,i}$ represents the $i$-th column of the matrix $\vb{T}$.
In a Monte Carlo inspired view, the process in Eq.~\eqref{eq:spawning} can be seen as each entry in $\vb{c}$ independently spawning the entries of the vector $\vb{T}_{:,i}\,c_i$. These spawns are then collected and summed together. This summing is often referred to as walker annihilation (when contributions of different signs are added) and is a critical step that allows FCIQMC to overcome the sign problem through cancellation of positive and negative contributions to a vector~\cite{boothFermionMonteCarlo2009}.

We convert the matrix $\vb{T}$ to a stochastically sampled operator $\check{\vb{T}}$ in two distinct phases, the first of which is matrix compression. Matrix compression replaces the spawning step from Eq.~\eqref{eq:spawning} with a stochastic procedure that only makes use of some elements of the original operator $\vb{T}$.
Suppose we are spawning the column corresponding to the coefficient $c_i$, and that the column $\vb{T}_{:,i}$ contains $m_i$ non-zero off-diagonal elements. We start by determining the number of spawning attempts from address $i$
\begin{equation}
  n_i = \lceil\abs{c_i}\rceil\,.
\end{equation}
where $ \lceil x\rceil$ denotes the smallest integer greater than or equal to $x$.
Next, we pick $n_i$ non-zero off-diagonal elements from the column uniformly at random with replacement.
Then we obtain the matrix-compressed operator, denoted as $\tilde{\vb{T}}$, by constructing each of its stochastically compressed columns
\begin{equation}
  \tilde{\vb{T}}_{:,i} \equiv \sum_j \tilde{T}_{j,i} \,\vb{j}
\end{equation}
where
\begin{equation}\label{eq:column-compression}
  \tilde{T}_{j,i} =
  \begin{cases}
    T_{i,i} & \text{if $j = i$,}\\
    \frac{m_i}{n_i} T_{j,i} & \text{if $j$ was selected in the picking step,}\\
    0 & \text{otherwise.}
  \end{cases}
\end{equation}
If the requested number of spawns is greater than $m_i$, we set $\tilde{\vb{T}}_{:,i} = \vb{T}_{:,i}$, \ie no stochastic compression takes place and the spawning step is executed exactly. This results in a semistochastic algorithm similar to the one described in Ref.~\cite{petruzieloSemistochasticProjectorMonte2012a} and is discussed in more detail in Sec.~\ref{ssec:semistochastic}.
The factor of $m_i/n_i$ in Eq.~\eqref{eq:column-compression} represents the inverse probability of picking the element. The selected element is scaled by this factor so that $\E[\tilde{T}_{j,i}] = T_{j,i}$ holds for all matrix elements. Note that this procedure always leaves the diagonal element $T_{i,i}$ unchanged, and if the magnitude of the coefficient $|c_i|$ is large enough, the column remains uncompressed.

\begin{figure}[t!]
\includegraphics[width=\textwidth]{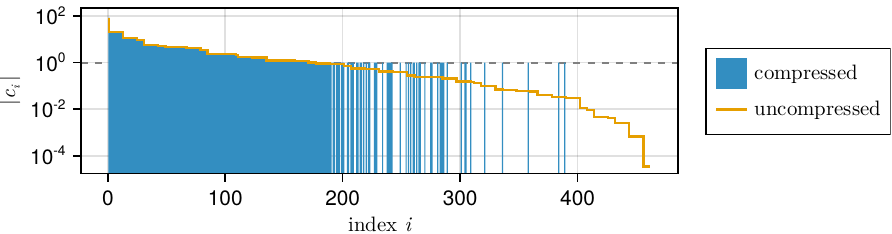}
    \caption{\label{fig:vec-compression}Stochastic vector compression. The plot shows the values $c_i$ at a given index $i$ of an uncompressed vector (yellow line) with a stochastically compressed vector (blue fill) superimposed on it. In the compressed vector, the coefficients with value $c_i \ge 1$ are unaffected by compression, while those below 1 are stochastically set to either 0 or 1 with probability proportional to $c_i$.
}
\end{figure}

The other building block in the compression scheme is stochastic vector compression of the coefficient vector. The goal here is to reduce the number of non-zero values in the coefficient vector $\vb{c}$ without changing the vector in expectation. This is done by applying a stochastic projection scheme $\check{\vb{\Pi}}$, which is defined as follows:
\begin{equation}\label{eq:vec-compression}
  {\left(\check{\vb{\Pi}}\vb{c}\right)}_i =
  \begin{cases}
    c_i & \text{if $\abs{c_i} \ge 1$},\\
    1 & \text{if $\abs{c_i} < 1$ with probability $\abs{c_i}$},\\
    0 & \text{if $\abs{c_i} < 1$ with probability $1 - \abs{c_i}$.}
  \end{cases}
\end{equation}
Similarly to the matrix compression scheme described above, this procedure has the property that increasing the 1-norm of the vector $\vb{c}$ lessens the effect $\check{\vb{\Pi}}$ has on it. A plot showing an example of a vector compressed this way is shown in Fig.~\ref{fig:vec-compression}. The figure shows the ground state eigenvector of the real space Bose-Hubbard Hamiltonian with $N=6$ particles in $M=6$ sites with an interaction strength of $u/t = 6$, which was normalized such that $\Norm{\vb{c}} = 1000$.

Finally, we combine matrix compression and vector compression to realise the stochastically compressed operator $\check{\vb{T}}$, defined by
\begin{equation}
  \check{\vb{T}}\vb{c} = \check{\vb{\Pi}}\left(\sum_i\tilde{\vb{T}}_{:,i}c_i\right)\,.
\end{equation}
Using the compressed operator $\check{\vb{T}}$ in place of $\vb{T}$ reduces the computational cost of a single matrix-vector multiplication considerably. When there is a small region in the vector where the coefficients $c_i$ are high, with the rest of the vector very close to zero, the vector compression scheme leaves the localised region mostly intact, while heavily compressing the (less important) low-valued regions. Matrix compression also ensures that the spawns from important, high-valued regions of the vector are performed essentially exactly, while the spawns from the low-valued regions of the vector are performed in a noisy manner.
A consequence of this is that FCIQMC works best when the ground state eigenvector of $\vb{H}$ is strongly localised. Where possible, the basis should be chosen so that the ground state is localised in Fock space.

When using the stochastic algorithm, the target norm $N_\mathrm{t}$ in Eq.~\eqref{eq:shift-update} plays an important role. By increasing $N_\mathrm{t}$, the stochastic operation $\check{\vb{T}}^{(n)}\vb{c}^{(n)}$ tends towards its exact counterpart, $\vb{T}^{(n)}\vb{c}^{(n)}$. Thus, simulations with higher $N_\mathrm{t}$ will be less noisy, giving better eigenvalue estimates, but be computationally more intensive. Setting a higher $N_\mathrm{t}$ also helps mitigate two issues present in stochastic FCIQMC, namely the population control bias (see Sec.~\ref{sec:pcb} below) and the sign problem.
The sign problem presents itself in Hamiltonians with positive off-diagonal elements. When the compressed vector representation is too small to fully capture the sign structure of the true eigenvector, FCIQMC converges to the wrong eigenpair~\cite{spencerSignProblemPopulation2012}. It can be overcome by using a $N_\mathrm{t}$ above a critical value $N_\mathrm{c}$, or be traded for an (often small) bias with the initiator approximation~\cite{clelandCommunicationsSurvivalFittest2010}.

Summarising, the FCIQMC algorithm is a stochastic process that is defined by the following update equation for the coefficient vector
\begin{align} \label{eq:fciqmc-stochastic}
      \vb{c}^{(n + 1)} = \check{\vb{T}}^{(n)} \vb{c}^{(n)}\,.
\end{align}
Compared to the deterministic iteration of Eq.~\eqref{eq:fciqmc}, now the application of the transition operator
\begin{align} \label{eq:stochasticTransitionOperator}
      \check{\vb{T}}^{(n)} = \check{\vb{\Pi}}\left[\vb{1} + \dt\ \left(S^{(n)}\vb{1} - \tilde{\vb{H}}\right)\right]\,,
\end{align}
includes stochastic compression of the Hamiltonian $\tilde{\vb{H}}$ and a stochastic vector compression $\check{\vb{\Pi}}$ of the final vector, as explained above. The stochastic vector-update Eq.~\eqref{eq:fciqmc-stochastic} is combined with a deterministic update equation for the shift parameter $S^{(n)}$ in order to avoid exponential growth or decay. This is done by default in each step using Eq.~\eqref{eq:shift-update}, corresponding to the (default) shift-update strategy \inline{DoubleLogUpdate()}.

\subsection{The population control bias} \label{sec:pcb}

The population control bias is an unavoidable stochastic bias in projector Monte Carlo methods that originates from controlling the norm of the coefficient vector with a fluctuating shift parameter in Eqs.~\eqref{eq:fciqmc-stochastic} and \eqref{eq:stochasticTransitionOperator}. The problem arises from the fact that fluctuations in the shift $S^{(n)}$ and the coefficient vector $\vb{c}^{(n)}$ are correlated, and thus the long-time average of Eq.~\eqref{eq:fciqmc-stochastic} is not simply the fixed point of the deterministic Eq.~\eqref{eq:fciqmc}. The result is that the average value of the shift is increased~\cite{brandStochasticDifferentialEquation2022} and as a secondary consequence all observables will obtain a small bias. It was shown in Ref.~\cite{brandStochasticDifferentialEquation2022} how a scalar model with noise can illustrate the origin of the population control bias. As this bias originates in the stochastic fluctuations of the shift and the coefficient vector, the best strategy to mitigate the population control bias is to minimise the noise.

The primary variable that controls the amount of noise in an FCIQMC calculation is the target norm $N_\mathrm{t}$, or walker number. Increasing the walker number will reduce the noise in the calculation. A generic argument shows that the population control bias scales inversely proportional to the walker number (assuming a constant threshold of stochastic compression) \cite{umrigarDiffusionMonteCarlo1993,ghanemPopulationControlBias2021}, although slower transient decay of the bias with walker number has been observed in some cases \cite{brandStochasticDifferentialEquation2022}. Besides increasing the walker number (which implies increased computational resources), a good strategy to reduce the stochastic noise is importance sampling \cite{umrigarDiffusionMonteCarlo1993,ghanemPopulationControlBias2021}. Importance sampling in the context of FCIQMC is a similarity transformation of the Hamiltonian with a diagonal matrix that is determined by a guiding function that approximates the desired solution for the coefficient vector \cite{ghanemPopulationControlBias2021}. It is available in Rimu.jl through a number of predefined wrapper types, e.g.~\inline{GutzwillerSampling} and \inline{GuidingVectorSampling}. These can easily be extended to custom guiding functions using the \inline{ModifiedHamiltonian} mechanism and interface, as described in Sec.~\ref{ssec:ham-wrappers}, or through the use of the \href{github.com/RimuQMC/Gutzwiller.jl}{Gutzwiller.jl} package.

A method to undo the population control bias by postprocessing FCIQMC data known as reweighting will be discussed in Sec.~\ref{sec:reweighting}.

\section{Computing observables}
\label{sec:Observables}

The main output of an FCIQMC simulation in Rimu.jl are the time series of various quantities.
By default, at each step the algorithm will record the value of the shift $S^{(n)}$, the vector norm $\Norm{\vb{c}^{(n)}}$ and several diagnostic quantities.
Additional observables can be calculated during the simulation in which case time series of vector-vector and vector-operator-vector products will also be reported.
The \inline{StatsTools} module in Rimu.jl contains a range of functions for calculating observables from the time series after an FCIQMC simulation.
Many of the convenience functions in the \inline{StatsTools} module are specific to analysis for the FCIQMC algorithm and the typical Hamiltonian models studied,
but the underlying functionality is flexible and extensible.

For this section, we will refer to a simple problem using the one-dimensional Bose-Hubbard Hamiltonian of 6 particles in 6 lattice sites.
\begin{jlcodeblock}
initial_address = near_uniform(BoseFS{6,6})
H = HubbardRealSpace(initial_address; u = 6.0)
\end{jlcodeblock}
The implementation of Fock addresses and Hamiltonians is explained in Section \ref{sec:implementation}.
For now, we perform the simulation for 3000 time steps.
\begin{jlcodeblock}
last_step = 3000;
problem = ProjectorMonteCarloProblem(H; last_step)
df = DataFrame(solve(problem))
# Progress: 100%|| Time: 0:00:00
# 3000×9 DataFrame
#   Row │ step   len    shift       norm       exact_steps  inexact_steps  ⋯
#       │ Int64  Int64  Float64     Float64    Int64        Int64          ⋯
# ──────┼───────────────────────────────────────────────────────────────────
#     1 │     1      2  -0.0409038    11.0               0              1  ⋯
#     2 │     2      3  -0.100568     12.1052            0              2
#     3 │     3      3   0.53137      12.216             0              3
#     4 │     4      4   0.1822       13.9003            0              3
#     5 │     5      4   0.581651     14.3938            0              4  ⋯
#     6 │     6      6  -0.0717713    16.9459            0              4
#   ⋮   │   ⋮      ⋮        ⋮           ⋮           ⋮             ⋮        ⋱
#  2995 │  2995    245  -4.0407     1000.88             25            214
#  2996 │  2996    245  -4.03679    1000.39             25            220
#  2997 │  2997    238  -3.98848     994.476            25            220  ⋯
#  2998 │  2998    232  -3.95599     990.631            24            214
#  2999 │  2999    233  -3.97726     993.4              25            207
#  3000 │  3000    233  -3.98744     994.77             25            208
#                                            3 columns and 2988 rows omitted
\end{jlcodeblock}
The output \inline{df} is a \inline{DataFrame}~\cite{bouchet-valatDataFramesjlFlexibleFast2023} whose columns are the time series of the quantities produced during the FCIQMC simulation.
In the following sections, we will describe how the time series are analysed to produce physically meaningful observables.

\subsection{Decorrelating a time series with blocking transformations}
\label{sec:Timeseries}

Even after equilibration, consecutive time steps of a QMC simulation are correlated, and the data must be decorrelated to obtain accurate estimates.
For a general stationary time series quantity $X$, assuming ergodicity, the expected value $\langle X \rangle$ is the long-time limit of the sample mean
\begin{align}
  \langle X \rangle &= \lim_{\Omega \to \infty} \overline{X} ,
\end{align}
where
\begin{align}
  \overline{X} = \frac{1}{\Omega} \sum_{n=1}^\Omega X^{(n)} ,
  \label{eq:SampleMean}
\end{align}
is the sample mean of a time series with $\Omega$ time steps.
A quantity of primary interest is the variance of the sample mean defined by
\begin{align}
  \sigma_{\overline{X}}^2 = \langle \overline{X}^2 \rangle -  \langle \overline{X} \rangle^2 ,
\end{align}
and the related standard error $\sigma_{\overline{X}}$.
This standard error quantifies the statistical uncertainty we have about the position of the expected value, given a sample mean calculated from a finite time series.
Estimating the variance $\sigma_{\overline{X}}^2$ from a given time series is complicated by the fact that nearby elements of the time series will typically be strongly correlated in a QMC calculation, \eg due to the fact that only a small step is performed in each iteration.

The main method in Rimu.jl to remove correlations from time series uses the so-called blocking analysis~\cite{flyvbjergErrorEstimatesAverages1989b}.
The variance of the sample mean can be expressed as
\begin{equation}
    \sigma_{\overline{X}}^2 = \frac{\gamma_0}{\Omega} + \frac{2}{\Omega} \sum_{t=1}^{\Omega-1} \left( 1 - \frac{t}{\Omega} \right) \gamma_t,
    \label{eq:VarianceCorrelationFunctions}
\end{equation}
where $\gamma_t = \langle X^{(i)} X^{(j)} \rangle - \langle X^{(i)} \rangle \langle X^{(j)} \rangle$ expresses the (unknown) correlation function of the time series, which is expected to only depend on the difference $t = |i - j|$.
The blocking analysis works by transforming the time series iteratively, and reducing its correlations, by re-blocking the data.
At each iteration a new time series $X'$ is built from `blocks' which are the average of consecutive steps
\begin{equation}
  X'^{(i)} = \frac{1}{2} \left( X^{(2i-1)} + X^{(2i)} \right).
  \label{eq:BlockingTransformation}
\end{equation}
The new time series has length $\Omega' = \Omega/2$ while the sample mean $\overline{X'} = \overline{X}$ and variance $\sigma_{\overline{X'}}^2 = \sigma_{\overline{X}}^2$ are unchanged.
However, the correlation functions are altered (and reduced) by the transformation Eq.~\eqref{eq:BlockingTransformation} such that at each iteration the leading-order term $\gamma'_0/\Omega'$ increases while the higher-order terms $t > 0$ decrease.
After some number of iterations the size of the blocks will exceed the correlation time scale of the original data and the blocks will be uncorrelated.
At this point the leading-order term in Eq.~\eqref{eq:VarianceCorrelationFunctions} will plateau (within random fluctuations) and higher-order terms will be negligible so that $[\tilde{\sigma}(k)]^2 \equiv {(\overline{X'^2} - \overline{X'}^2)/\Omega'}\approx {\gamma_0'/\Omega'}$
is a good estimate and evaluating the unknown correlation function at finite time can be avoided.
The blocking analysis also provides an estimate of the uncertainty in $\sigma_X^2$ and in practice this uncertainty in $\sigma_X^2$ rises after the estimates for $\sigma_X^2$ have plateaued.
The detection of this plateau and the minimum number of blocking iterations to achieve it are automated with the $M$-test \cite{jonssonStandardErrorEstimation2018}, which detects when the higher-order correlations, in particular $\gamma_1$, are indistinguishable from zero at a specified confidence level.

\begin{figure}[t!]
\includegraphics[width=\textwidth]{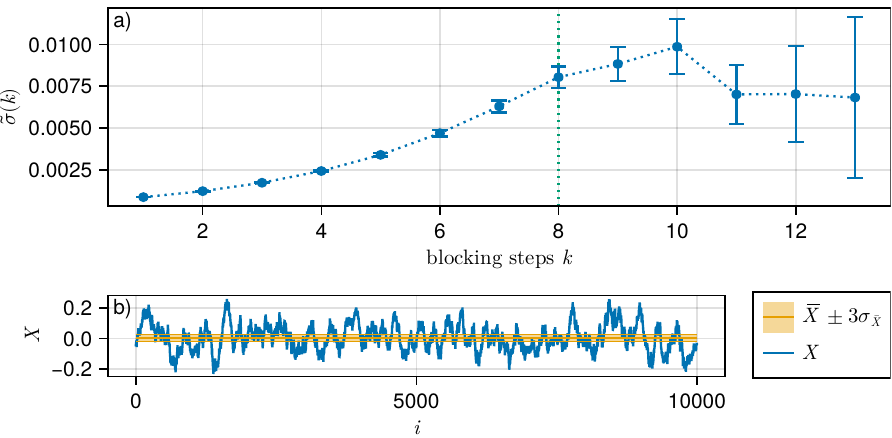}
\caption{\label{fig:m-test}The blocking analysis with $M$-test on a correlated time series of synthetic data $X$ to estimate the root variance of the sample mean $\sigma_{\bar{X}}$, i.e.\ the standard error of $\bar{X}$. Panel a) shows how the na\"ive  standard error (assuming uncorrelated data) $\tilde{\sigma}(k) \equiv \sqrt{(\overline{X'^2} - \overline{X'}^2)/\Omega'}$ after $k-1$ blocking transformations changes as the number of steps increases. Note that past $k=8$, the values of $\tilde{\sigma}(k)$ are equivalent up to  their error bars, hence the $M$-test identifies $k=8$ as the appropriate level of blocking and $\sigma_{\bar{X}} \approx \tilde{\sigma}(k=8)$ is chosen as the best estimate for the standard error. Panel b) shows the raw data with the calculated mean and estimated standard error superimposed on it.}
\end{figure}

In Rimu.jl the blocking analysis is provided by the \inline{blocking_analysis} function in the \inline{StatsTools} module that can be applied to a time series:
\begin{jlcodeblock}
correlated_ts = smoothen(randn(10_000), 128)

blocking_analysis(correlated_ts)
# BlockingResult{Float64}
#   mean = -0.0027 ± 0.0095
#   with uncertainty of ± 0.0007686451776019342
#   from 78 blocks after 7 transformations (k = 8).
\end{jlcodeblock}
See Fig.~\ref{fig:m-test} for a graphical depiction of the blocking analysis with $M$-test performed on this data. The per-step blocking data shown in Fig.~\ref{fig:m-test}a) is conveniently obtained using \inline{blocking_analysis_data(correlated_ts)}.

\subsection{Energy estimators}
\label{sec:EnergyEstimators}

The simplest observable that can be calculated in an FCIQMC calculation is the expectation value of the shift $\langle S \rangle$, which is an estimator of the ground state energy.
The update procedure Eq.~\eqref{eq:shift-update} means that the shift behaves as a critically damped harmonic oscillator on average.
After the initial equilibration phase $S^{(n)}$ is a stationary time series fluctuating about the expected value $\langle S \rangle$.
Blocking analysis is sufficient to decorrelate the data and obtain an energy estimator using the shift time series $S^{(n)}$.
This is the most common analysis for FCIQMC so \inline{StatsTools} has a convenience function called \inline{shift_estimator} for performing the blocking analysis on the output of a simulation.
This function automatically extracts the correct column from the output \inline{DataFrame} and performs a blocking analysis on the time series.
For the output \inline{df} of solving the \inline{ProjectorMonteCarlo} from the start of this section we have
\begin{jlcodeblock}
shift_estimator(df; skip = 1_000)
# BlockingResult{Float64}
#   mean = -4.0225 ± 0.0034
#   with uncertainty of ± 0.00021708770311760708
#   from 125 blocks after 4 transformations (k = 5).
\end{jlcodeblock}
Note the keyword argument to skip the first steps and avoid the equilibration phase.

While the shift estimator $\langle S \rangle$ is the easiest to calculate, there are other useful estimators of the ground state energy, such as the projected energy
\begin{equation}
  \overline{E}_{\vb{y}} = \frac{\langle \vb{y}^\dagger \vb{H} \vb{c} \rangle}{\langle \vb{y}^\dagger \vb{c} \rangle},
  \label{eq:ProjectedEnergy}
\end{equation}
where $\vb{y}$ is a fixed reference vector.
Different common energy estimators can be understood as a projected energy with different choices for the reference vector $\vb{y}$. This is discussed in detail in Ref.~\cite{brandStochasticDifferentialEquation2022}, where also estimates for the effect of the population control bias for the various estimators are provided.
Of practical interest is the choice of a fixed vector $\vb{y}$ containing only a small number of (possibly only a single) nonzero coefficients, as this makes the computation efficient.

The estimator in Eq.~\eqref{eq:ProjectedEnergy} is obtained in two steps.
First, for each time step, after spawning and annihilation is finished, the instantaneous vector $\vb{c}^{(n)}$ and the Hamiltonian are used to separately calculate $\vb{y}^\dagger \vb{H} \vb{c}^{(n)}$ and $\vb{y}^\dagger \vb{H} \vb{c}^{(n)}$, the constituents of the numerator and denominator of Eq.~\eqref{eq:ProjectedEnergy}, respectively. Both are recorded as time series and reported to the \inline{DataFrame} of the simulation.
After the simulation is finished, the two
time series should be combined to obtain $\overline{E}_{\vb{y}}$.
This estimator is therefore a ratio of fluctuating quantities and propagation of uncertainty must be performed carefully, particularly when the denominator may be near zero due to fluctuations.

For a quantity of interest $\langle X/Y \rangle$ defined by two time series $X$ and $Y$ the sample means $\overline{X}$ and $\overline{Y}$ define the mean value of the ratio $r = \overline{X}/\overline{Y}$. As we are dealing with correlated time series, the error $\sigma_r$ on this ratio requires estimating the errors $\sigma_X$ and $\sigma_Y$ by performing the blocking analysis separately on each time series. These values and the covariance $\rho = \mathrm{cov}(X,Y)$ define a correlated normal distribution of the ratio $r$. This distribution must be sampled to determine the error $\sigma_r$ on the ratio of means $r = \overline{X}/\overline{Y}$.
In Rimu.jl this sampling relies on Monte Carlo error propagation provided by the package \href{https://github.com/baggepinnen/MonteCarloMeasurements.jl}{MonteCarloMeasurements.jl}~\cite{carlsonMonteCarloMeasurementsjlNonlinearPropagation2020a} .
The calculation of ratio estimators is provided by the \inline{ratio_of_means} function in the \inline{StatsTools} module.
\begin{jlcodeblock}
Xs = 2 .+ smoothen(randn(10_000), 128)
Ys = 1 .+ smoothen(randn(10_000), 64)
ratio_of_means(Xs, Ys)
# RatioBlockingResult{Float64,MonteCarloMeasurements.Particles{Float64, 2000}}
#   ratio = 1.96085 ± (0.0205071, 0.0179696) (MC)
#   95% confidence interval: [1.92504, 2.00045] (MC)
#   linear error propagation: 1.96205 ± 0.0193621
#   |δ_y| = |0.00868362| (≤ 0.1 for normal approx)
#   Blocking successful with 78 blocks after 7 transformations (k = 8).
\end{jlcodeblock}
The output is a \inline{RatioBlockingResult} which contains information about the ratio distribution.
In order for this method to work, the ratio distribution should be normally distributed, which is a good approximation if the parameter $\delta_Y = \sigma_Y/\overline{Y}$ is small (typically $\lesssim 0.1$).

To calculate a projected energy during an FCIQMC simulation, first define a post-step strategy which indicates how to use existing information to calculate additional quantities when solved.
For example, we could choose the fixed reference vector to be the initial vector $\vb{y} = (N_w,0,\ldots)$, where $N_w$ is the initial number of walkers at the start of the simulation, (by default $N_w = 10$). That is, all walkers start on the initial Fock address, and all other Fock addresses have zero walkers initially.
\begin{jlcodeblock}
initial_vector = default_starting_vector(H)
post_step_strategy = ProjectedEnergy(H, initial_vector)
\end{jlcodeblock}
Then the simulation is re-run with this post-step strategy
\begin{jlcodeblock}
problem = ProjectorMonteCarloProblem(H; last_step, post_step_strategy)
df = DataFrame(solve(problem))
\end{jlcodeblock}
The \inline{DataFrame} \inline{df} now contains additional columns \inline{"hproj"} and \inline{"vproj"} containing the time series for the numerator and denominator, respectively, of the projected energy.
The \inline{StatsTools} module provides a convenience function \inline{projected_energy} that automatically extracts the correct time series from the output and calculates the ratio of means.
\begin{jlcodeblock}
projected_energy(df; skip = 1_000)
# RatioBlockingResult{Float64,MonteCarloMeasurements.Particles{Float64, 2000}}
#   ratio = -4.02064 ± (0.00195644, 0.00216691) (MC)
#   95% confidence interval: [-4.02469, -4.01672] (MC)
#   linear error propagation: -4.02066 ± 0.00205069
#   |δ_y| = |0.000765156| (≤ 0.1 for normal approx)
#   Blocking successful with 62 blocks after 5 transformations (k = 6).
\end{jlcodeblock}
The \inline{StatsTools} module provides other convenience functions for energy estimators such as \inline{variational_energy_estimator} for calculating the variational energy, but this estimator requires that the simulation is run with more than one replica, explained in the next section.

\subsection{General observables}
\label{sec:GeneralObservables}

In addition to estimators of the ground state energy, Rimu.jl is able to calculate general observables.
Since the instantaneous vectors are unnormalised we wish to calculate expected values of a general operator $\hat{Q}$ as a Rayleigh quotient
\begin{equation}
  \langle \hat{Q} \rangle = \frac{\langle \mathbf{c}^\dagger \vb{Q} \mathbf{c} \rangle}{\langle \mathbf{c}^\dagger\mathbf{c} \rangle}.
  \label{eq:RayleighQuotient}
\end{equation}
The implementation of the matrix representation $\vb{Q}$ will be discussed in Section \ref{ssec:hamiltonians}.
The Rayleigh quotient also depends on the state vector $\vb{c}$ which will, of course, be approximated during a Monte Carlo simulation.
However, even keeping an estimate of the entire (\ie uncompressed) vector can be computationally prohibitive when the Hilbert space scales exponentially.
Instead, here we discuss how the numerator and denominator of Eq.~\eqref{eq:RayleighQuotient} are computed via the replica trick, using multiple independent simulations. For further details and demonstration of the effectiveness of this technique see Ref.~\cite{brandStochasticDifferentialEquation2022}.

In Rimu.jl an FCIQMC simulation can be run with $R$ instances, or replicas, to be run simultaneously and independently.
At each time step, the FCIQMC algorithm propagates vectors $\vb{c}_A$, $A = 1,\ldots,R$.
They are statistically independent in that $\mathrm{cov}(\vb{c}_A, \vb{c}_B) = 0$ but model the same system \ie~$\langle \vb{c}_A \rangle = \langle \vb{c}_B \rangle = \vb{v}_0$.
For each pair of replicas $(A,B)$ we calculate the vector-vector dot-product $\vb{c}_A^{(n)\dagger} \vb{c}_B^{(n)}$ and vector-operator-vector dot-product $\vb{c}_A^{(n)\dagger} \vb{Q} \vb{c}_B^{(n)}$.
These products are reported as time series so that the expected value $\langle \hat{Q} \rangle$ is then estimated as
\begin{equation}
  \overline{Q} = \frac{ \sum_{(A,B)}^R \overline{\vb{c}_A^\dagger \vb{Q} \vb{c}_B} }{\sum_{(A,B)}^R \overline{\vb{c}_A^\dagger \vb{c}_B}},
  \label{eq:RayleighQuotientReplica}
\end{equation}
where the summations go over pairs of replicas with $A<B$. The sample means $\overline{\cdot}$ in the numerator and denominator are obtained with the blocking analysis discussed in Section \ref{sec:Timeseries} and the overall estimator for $\overline{Q}$ is obtained with the ratio estimator discussed in Section \ref{sec:EnergyEstimators}.

From Eq.~\eqref{eq:RayleighQuotientReplica} it is necessary to use at least two replicas and it can be beneficial to use more.
Due to random fluctuations and sparseness of the vectors $\vb{c}_A$, the vector-vector product for any one pair may be sampled close to zero.
Using more than two replicas reduces the variance in the overall denominator of Eq.~\eqref{eq:RayleighQuotientReplica}, which scales as $1/\sqrt{R (R - 1) \Omega}$, and thus the ratio estimator $\overline{Q}$ also has smaller variance.
However, the computational cost of $R$ replicas each running for $\Omega$ time steps scales as $R\Omega$ and the cost of computing dot products between replicas scales as $R^2$.
In practice, we find that $R = 2,3 \text{ or } 4$ is sufficient in most cases for greatly improving the statistics of general observables without ballooning computational cost.

Calculating the expectation values of observables was already demonstrated for the density-density correlation function for the Bose-Hubbard model in the code example in Sec.~\ref{sec:example-observables}. As demonstrated there, the unbiased estimator of Eq.~\eqref{eq:RayleighQuotientReplica} can be conveniently calculated using the function \inline{rayleigh_replica_estimator}, which performs the appropriate averaging over the replicas and provides an error estimate using blocking analysis and Monte Carlo error propagation. The results of the calculation are shown in Fig.~\ref{fig:g2-plot}.

An observable expectation value of particular interest is the variational energy, denoted $\bar{H}$, which is defined by using the system Hamiltonian as an observable in Eq.~\eqref{eq:RayleighQuotientReplica}. The variational energy has the advantage that it is less affected by the population control bias than the shift, while at the same time providing an upper bound on the exact energy of the system \cite{brandStochasticDifferentialEquation2022}. While it could be calculated using the general procedure for observables described above, i.e.~by passing the Hamiltonian into \inline{AllOverlaps}, there is a more efficient way that avoids computing dot products with the Hamiltonian operator.
Instead, the variational energy is obtained by using the average of the shifts from different replicas, computing the numerator in Eq.~\eqref{eq:RayleighQuotientReplica} for each pair of replicas as $\overline{\vb{c}_A^\dagger \vb{Q} \vb{c}_B}  = \tfrac{1}{2}\overline{( S_A + S_B ) \vb{c}_A^\dagger \vb{c}_B}$ \cite{brandStochasticDifferentialEquation2022}.
The convenience function \inline{variational_energy_estimator} combines the shift time series from all replicas to produce a ratio of means.
\begin{jlcodeblock}
problem = ProjectorMonteCarloProblem(H; last_step, n_replicas = 2)
df = DataFrame(solve(problem))
variational_energy_estimator(df; skip = 1_000)
# RatioBlockingResult{Float64,MonteCarloMeasurements.Particles{Float64, 2000}}
#   ratio = -4.02149 ± (0.00248037, 0.00245498) (MC)
#   95% confidence interval: [-4.02635, -4.01664]) (MC)
#   linear error propagation: -4.02163 ± 0.00248588
#   |δ_y| = |0.000943422| (≤ 0.1 for normal approx)
#   Blocking successful with 62 blocks after 5 transformations (k = 6).
\end{jlcodeblock}
Multiple replicas are required for using \inline{variational_energy_estimator}, otherwise the function will error.

\subsection{Comparison of energy estimators}
\label{sec:estimator-comparison}
Because this example is for a small system, the three energy estimators considered so far obtain similar estimates of the ground state energy $E = -4.02$.
For more realistic problems where the Hilbert space is much larger than the number of walkers and the population control bias can be significant, it is useful to have different estimators for estimating the ground state energy. The mean of the shift has the benefit of being always available if the calculation converges at all. However it has the largest population control bias \cite{brandStochasticDifferentialEquation2022}, see Sec.~\ref{sec:pcb}.
The variational energy estimator is nominally better as it has a reduced bias. However, it requires multiple replicas and can become very noisy if the overlap between different replicas becomes very small. This can occur (typically for sign-problem-free calculations) in situations where the state vectors are spread out thinly in Hilbert space with coefficient values near the stochastic projection threshold (by default set to 1).
The projected energy estimator is particularly useful when the ground state wave function is compact and dominated by only a few large coefficients (and a good reference vector is available).
While the population control bias is proven to be smaller than that of the shift, the absence of a variational bound means that the effect of the bias could shift the projected energy to higher or lower values compared to the exact energy.
For a given problem it may require experimenting to find out which energy estimator provides the smallest statistical error.


\subsection{Reweighting} \label{sec:reweighting}

Finally, we briefly mention an advanced technique for analysing the time series produced by FCIQMC simulations.
As mentioned in Sec.~\ref{ssec:fciqmc:montecarlo}, the population control bias is always present when the size of the Hilbert space is much larger than the number of walkers.
Methods for dealing with the population control bias in FCIQMC are discussed in detail in Ref.~\cite{brandStochasticDifferentialEquation2022}, including a procedure called reweighting, which is based on the methods of Ref.~\cite{umrigarDiffusionMonteCarlo1993}.
The idea is to consider an idealised FCIQMC procedure where the coefficient vector is altered by a set of weights in such a way as to counter the effect of fluctuations in the shift $S^{(n)}$, thus reducing the population control bias.
Importantly, the reweighting procedure is not actually applied to the coefficient vector $\vb{c}^{(n)}$ during the simulation and instead the output time series of an ordinary FCIQMC simulation are adjusted by weights $w_h^{(n)}$ over a window with size $h$.
In principle, energy estimators can be constructed from reweighted time series that are asymptotically unbiased as $h \to \infty$ and $\dt \to 0$.
In practice the variance grows as $h$ becomes large, but for small $h$ there can be an improvement in the population control bias.
The efficacy of the reweighting procedure for FCIQMC is explored more thoroughly in Ref.~\cite{brandStochasticDifferentialEquation2022}.
Here we remark that the \inline{StatsTools} module contains convenience functions \inline{mixed_estimator} and \inline{growth_estimator} to calculate energy estimators using the reweighting technique, and \inline{rayleigh_replica_estimator} can also use reweighting to reduce the effect of the population control bias on estimates of observables.

\section{Exact diagonalisation}\label{sec:exact-diagonalization}

While the main focus of Rimu.jl is projector quantum Monte Carlo, it also supports exact diagonalisation through a convenient interface with the \inline{ExactDiagonalizationProblem} type.
The diagonalisation can be performed in two ways --- through explicitly building a sparse matrix, and through matrix-free methods.
The diagonalisation then proceeds using algorithms from the Julia package ecosystem including \href{https://github.com/Jutho/KrylovKit.jl}{KrylovKit.jl}~\cite{haegemanKrylovKit2025}, \href{https://github.com/JuliaLinearAlgebra/Arpack.jl}{Arpack.jl}~\cite{lehoucqARPACKUsersGuide1998}, and \href{https://github.com/JuliaLinearAlgebra/IterativeSolvers.jl}{IterativeSolvers.jl}.

The sparse matrix methods are straightforward; the Hamiltonian is converted to a matrix and passed to the solver. To convert an operator to a sparse matrix, we treat it as a graph whose vertices correspond to the basis elements and its edges correspond to non-zero off-diagonal elements of the operator. We then walk the graph in a parallel breadth-first search and record all encountered off-diagonal elements in a list that is later converted to a matrix. Converting a Hamiltonian to a matrix is only feasible with small problems, as the matrices quickly grow too large to store in memory, even if they are sparse.

For use with matrix-free methods, the Hamiltonian is wrapped in a way that follows the interface from \href{https://github.com/JuliaLinearAlgebra/LinearMaps.jl}{LinearMaps.jl}. This allows us to multiply the Hamiltonian with a standard Julia vector much like one would a \inline{Matrix}. An operator-vector product $\vb{w} = \vb{H} \vb{v}$ is then executed by calculating each element of $\vb{w}$ as
\begin{equation}
  w_i = \vb{H}_{i,:}\vb{v}\,,
\end{equation}
where $\vb{H}_{i,:}$ is the $i$-th row of $\vb{H}$.
This can be performed in parallel for each index $i$. Since in this case $i$ are integers and Rimu.jl operators are indexed by Fock addresses, a translation layer is used to convert between them, and the off-diagonal elements of $\vb{H}$ are generated on the fly as described in Sec.~\ref{ssec:hamiltonians}.
While these matrix-free methods are much slower than the methods employing sparse matrices, we only need to store the vectors and the translation layer in memory, which allows us to solve much larger problems.

\section{Data structures}\label{sec:implementation}

In Rimu.jl we refer to Fock state addresses, which are compact representations of many-body states (typically so-called Fock states, i.e.~states with a well-defined particle number) and form an orthonormal basis for the coefficient vectors $\vb c$. Because the basis is so large for typical many-body problems we need to store the coefficient vectors as sparse vectors that can be indexed by Fock addresses. Operators that act on the vectors, including Hamiltonians, are also indexed by Fock addresses but without storing the matrix elements in memory.
Efficient realisation of vector--vector and matrix--vector products depends on the implementation of these key data structures. In this section we discuss Fock state addresses in Sec.~\ref{ssec:addresses}, Hamiltonians and other operators in Sec.~\ref{ssec:hamiltonians} and vectors in Sec.~\ref{ssec:dvecs}. Finally, in \ref{ssec:parallel} we discuss how a matrix-vector product is parallelised and distributed.

For a more thorough treatment of the types and interfaces, as well as usage examples, please consult the \href{https://RimuQMC.github.io/Rimu.jl/stable/}{project documentation}.

\subsection{Fock state addresses}\label{ssec:addresses}

Rimu.jl is primarily designed to work with quantum many-body Hamiltonians. To represent them as matrices, we use a basis of Fock states. A Fock state for a single species of indistinguishable particles is denoted by its occupation number representation and defined as
\begin{equation}
  \ket{n_1,n_2,\dots,n_M} = \prod_{i=1}^M\frac{1}{\sqrt{n_i!}} {\left(\hat{a}_i^\dagger\right)}^{n_i}\ket{\text{vac}},
\end{equation}
where the $n_i$ are the occupation numbers, $\hat{a}_i^\dagger$ is the creation operator (for either a boson or a fermion, as appropriate), $N = \sum_{i=1}^M n_i$ is the number of particles, $M$ is the number of modes (sites) indexed by $i$, and $\ket{\text{vac}}$ is the vacuum state with no particles.
This state is represented in our code by a \inline{SingleComponentFockAddress} and
the following concrete types are currently available:
\begin{itemize}
\item \inline{BoseFS\{N,M\}}: Bosonic Fock states with $N$ particles in $M$ modes.
\item \inline{FermiFS\{N,M\}}: Fermionic Fock states with $N$ particles in $M$ modes. In the chemistry literature, fermionic Fock states are often referred to as `Slater determinants', even though for the purpose of indexing only the sequence of occupation numbers is relevant.
\item \inline{HardcoreBoseFS\{N,M\}}: Bosonic Fock state with at most one particle per mode. These states can represent spin-$\frac{1}{2}$ lattices with $M$ sites or qubit assemblies.
\end{itemize}
Specifying the particle number $N$ as a type parameter makes the particle number a part of the type and thus allows the Julia compiler to generate optimised code for storage and processing that is specific to this particle number. In order to treat models where the particle number is not conserved, Rimu.jl allows the type parameter for particle number to be set to \inline{missing}. In this case, particle-number-changing excitations can be performed on a Fock address without changing its type and the code remains type-stable, which is important for the generation of efficient compiled code.
Multi-species or spinor particle states can be represented by combining single component states into a \inline{CompositeFS}, to be discussed below.

Fock states are constructed by either passing the occupation number representation to the appropriate constructor:
\begin{jlcodeblock}
BoseFS(0, 4, 0) # BoseFS(0, 4, 0)
\end{jlcodeblock}
or by specifying the number of modes followed by (position $\Longrightarrow$ occupation number)-pairs:
\begin{jlcodeblock}
BoseFS(3, 2 => 4) == BoseFS(0, 4, 0) # true
\end{jlcodeblock}
In some cases, Fock states are printed in a compact string representation. This output can be copied back to the Julia REPL and used as a constructor:
\begin{jlcodeblock}
print(IOContext(stdout, :compact=>true), BoseFS(0, 4, 0)) # fs"|0 4 0⟩"
fs"|0 4 0⟩" # BoseFS(0, 4, 0)
\end{jlcodeblock}

Since we want to work with very large vectors, it is extremely important to have compact representations of bosonic and fermionic Fock states available. To achieve this, \inline{BoseFS}, \inline{FermiFS}, and \inline{HardcoreBoseFS} are represented either by bitstrings, sorted particle lists, or a vector of occupation numbers. Which representation is used is determined automatically at compile time, depending on which one is more compact or appropriate. The bitstrings are packed into unsigned integers of appropriate size, \eg a 9-bit bitstring would be stored in a 16-bit integer. If the number of bits is larger than 64, an appropriate number of 64-bit integers is used. The sorted particle list is a sorted sequence of integers, where each integer represents a particle, and its value represents the mode the particle is in. The width of the integers is determined automatically to allow for the specified number of modes $M$ and the list is kept sorted to ensure a one-to-one correspondence between the data structure and Fock states, which represent indistinguishable particles. Sorted particle lists are used only when $M \gg N$.

In the bitstring representation of \inline{BoseFS}, a sequence of ones is used to represent the number of bosons in a mode, and zeros are used as separators between modes. This representation requires $N + M - 1$ bits.
\begin{jlcodeblock}
bitstring(BoseFS(1, 0, 1)) # "1001"
bitstring(BoseFS(0,2,3,4,0,1)) # "100111101110110"
\end{jlcodeblock}
Note that the bitstring is stored in reversed order.

Fermionic addresses (or hard core bosons) behave in largely the same way as the bosonic ones, but do not allow for more than one particle in the same mode.
\begin{jlcodeblock}
FermiFS(1, 0, 1) # FermiFS(1, 0, 1)
FermiFS(3, 1 => 1, 3 => 1) # FermiFS(1, 0, 1)
FermiFS(2, 0, 1)
# ERROR: ArgumentError: Invalid ONR: may only contain 0s and 1s.
\end{jlcodeblock}
A \inline{FermiFS} bitstring is simple --- each bit in the bitstring represents a mode and its value indicates whether the mode is occupied or not. This requires $M$ bits to store the state regardless of particle number.
\begin{jlcodeblock}
bitstring(FermiFS(1, 0, 1)) # "101"
bitstring(FermiFS(1, 0, 1, 0, 0, 1, 1)) # "1100101"
\end{jlcodeblock}

The non-particle number-conserving \inline{BoseFS\{missing, M\}} uses a different representation. This type of address is represented by a sequence of $M$ integers, each representing the occupation number in a given mode. The width of these integers determines the maximum number of particles allowed per mode. It defaults to \inline{UInt8}, which allows for a maximum of 255 particles per mode.
\begin{jlcodeblock}
Int(maximum_mode_occupation(BoseFS{missing}(1, 2, 34))) # 255
Int(maximum_mode_occupation(BoseFS{missing}(1, 2, 256, 511; type=UInt16)))
# 65535
\end{jlcodeblock}

For more complicated models that require multiple particle species, such as mixtures or particles with spin, Rimu.jl provides
\begin{itemize}
    \item \inline{CompositeFS}: This container type can contain an arbitrary number and combination of bosonic or fermionic \inline{SingleComponentFockAddress}s and thus be used to represent spinor particle types or other multi-component Fock states.
\end{itemize}
A \inline{CompositeFS} can be constructed by passing other non-composite addresses.
\begin{jlcodeblock}
CompositeFS(FermiFS(1, 1, 0, 0), FermiFS(1, 0, 1, 0), BoseFS{missing}(2, 5))
# CompositeFS(
#   FermiFS(1, 1, 0, 0),
#   FermiFS(1, 0, 1, 0),
#   BoseFS{missing}(2, 5),
# )
\end{jlcodeblock}
Fock states with spin-$\frac{1}{2}$ fermions, where both spin-up (alpha) and spin-down (beta) states are present, are treated as a special case of \inline{CompositeFS} in Rimu.jl. This case is appropriate for quantum many-body problems with electrons. For convenience, a shorthand notation for defining such `Slater determinants' is available.
\begin{jlcodeblock}
FermiFS2C((1,0,1),(1,1,0))
# CompositeFS(
#   FermiFS(1, 0, 1),
#   FermiFS(1, 1, 0),
# )
\end{jlcodeblock}

To make writing operators that act on the Fock states more convenient, the \inline{excitation} function is provided. As an example, suppose that we want to apply the operator $\hat{a}^\dagger_3\hat{a}_5$
to the fermionic Fock state $\ket{1 0 0 1 1 0 1}$, ie. a particle in mode 5 is moved to mode 3. In code, this is done as follows. First, we define the state
\begin{jlcodeblock}
state = FermiFS(1, 0, 0, 1, 1, 0, 1)
\end{jlcodeblock}
Next, we use the \inline{find_mode} function to get pointers to the 3rd and 5th sites in the address.
\begin{jlcodeblock}
i, j = find_mode(state, (3, 5))
\end{jlcodeblock}
Finally, we apply the \inline{excitation} function, which takes as its arguments the state and the modes where particles are to be created and annihilated. The new state and the value associated with applying the operator are returned:
\begin{jlcodeblock}
excitation(state, (i,), (j,)) # (FermiFS(1, 0, 1, 1, 0, 0, 1), -1.0)
\end{jlcodeblock}
Using multiple dispatch, these tools can be used to write functions that will work on both fermionic and bosonic addresses and will take fermionic antisymmetry into account. Note that the \inline{excitation} function can be used for excitations that involve any number of particles.

\subsection{Hamiltonians and operators}\label{ssec:hamiltonians}

Conceptually, Hamiltonians can be represented as matrices of, possibly, enormous size with elements that are the overlap integrals of the Hamiltonian operator between the states of the Fock basis.
For problems where the state is represented by a very large vector, which may itself only represent a subset of an even larger Hilbert space, it is impractical to represent operators in the standard way using dense (or even sparse) matrices. Thus we resort to matrix-free, abstract representations of operators in Rimu.jl. In order to enable the use of such an operator $\hat{A}$ with matrix-free linear algebra routines~\cite{haegemanKrylovKit2025,lehoucqARPACKUsersGuide1998} and FCIQMC, only two operations are required: performing the operator-vector product $\hat{A}\ket{\Psi}$ and stochastically selecting an element from a matrix column with known probability.

Rimu.jl defines an interface for working with Hamiltonians and operators, which revolves around the concept of an operator column. This is implemented with the \inline{operator_column(A, j)} function, which, conceptually, represents
\begin{equation}
  \hat{A}\ket{j} = \sum_i\ket{i}\braket{i|\hat{A}|j}\,,
\end{equation}
where $j$ is the column address and $i$ represents all reachable addresses with non-zero matrix element $\braket{i|\hat{A}|j}$.
Stochastically sampling an element from an operator column is performed by the interface function \inline{random_offdiagonal}.
The full product of $\hat{A}$ with a state $\ket{\Psi}$ can be written as
\begin{equation}
  \hat{A}\ket{\Psi} = \sum_{j} \left(\sum_i\ket{i}\braket{i|\hat{A}|j}\right)\braket{j|\Psi}\,,
\end{equation}
which is a linear combination of operator columns.
Whether stochastically sampling the operator column or computing all of its elements, the required non-zero operator matrix elements $\braket{i|\hat{A}|j}$ can be evaluated efficiently on the fly without storing all of them in memory. This can reduce the memory requirements of exact diagonalisation routines by many orders of magnitude and allows us to study systems much larger than would otherwise have been possible.

When an operator is only intended to be used as an observable, generating the column is not required. Instead, Rimu.jl allows for the direct definition of the overlap
\begin{equation}
    \braket{\Phi|\hat{A}|\Psi}
\end{equation}
in terms of the Julia \inline{dot} function. To facilitate the different use-cases of operators in Rimu.jl, the following type hierarchy is in use:
\begin{jlcodeblock}
AbstractHamiltonian <: AbstractOperator <: AbstractObservable
\end{jlcodeblock}
The \inline{AbstractObservable} is the most permissive type and only supports vector-observable-vector dot-products. The next type, \inline{AbstractOperator}, permits additionally the generation of diagonal and off-diagonal elements from a given Fock state. Dot products, vector-operator products and conversion to Julia (sparse) matrices are implemented for \inline{AbstractOperator}. The last type, \inline{AbstractHamiltonian}, is used for model Hamiltonians and implements the interface of \inline{AbstractOperator}, as well as the \inline{starting\_address} function, which returns a Fock state address. This address is used to start an FCIQMC computation and defines the address type and the Hilbert (sub)space in which the operator is represented.
While we have presented in this section operators defined in terms of a Fock state basis, Rimu.jl is agnostic to what the basis of the linear space (i.e.\ Hilbert space) is. Addresses can in principle be of any type, and alternate bases like configuration state functions could be implemented by a user.

To demonstrate the concepts of Hamiltonians and operator columns in Rimu.jl, consider the 1D Bose-Hubbard model with strong interactions and 12 particles in 12 lattice sites (as seen in the example of Sec.~\ref{sec:example}).
\begin{jlcodeblock}
hamiltonian = HubbardRealSpace(near_uniform(BoseFS{12,12}); u=6)
\end{jlcodeblock}
While the \inline{hamiltonian} represents a matrix of size $1{,}352{,}078^2$, only 24 bytes are required to store it in memory:
\begin{jlcodeblock}
dimension(hamiltonian) # 1352078
sizeof(hamiltonian) # 24
\end{jlcodeblock}
The only things stored in the Hamiltonian object are the starting address, its parameters, and information about its geometry.

The matrix elements of the Hamiltonian can be accessed through the \inline{operator_column} function, which returns an object representing a matrix column indexed by a given address
\begin{jlcodeblock}
column = operator_column(hamiltonian, near_uniform(BoseFS{12,12}))
\end{jlcodeblock}
The column can then be used to extract the diagonal element
\begin{jlcodeblock}
diagonal_element(column) # 0.0
\end{jlcodeblock}
and a lazy iterator over the off-diagonal elements
\begin{jlcodeblock}
ods = offdiagonals(column)
# 24-element Rimu.Hamiltonians.HubbardRealSpaceColumnOffdiagonals{Float64, …
#  fs"|0 2 1 1 1 1 1 1 1 1 1 1⟩" => -1.4142135623730951
#  fs"|1 0 2 1 1 1 1 1 1 1 1 1⟩" => -1.4142135623730951
#  fs"|1 1 0 2 1 1 1 1 1 1 1 1⟩" => -1.4142135623730951
#  fs"|1 1 1 0 2 1 1 1 1 1 1 1⟩" => -1.4142135623730951
#                                ⋮
#  fs"|1 1 1 1 1 1 1 1 2 0 1 1⟩" => -1.4142135623730951
#  fs"|1 1 1 1 1 1 1 1 1 2 0 1⟩" => -1.4142135623730951
#  fs"|1 1 1 1 1 1 1 1 1 1 2 0⟩" => -1.4142135623730951
\end{jlcodeblock}
\inline{offdiagonals} here represents all the non-zero elements in the column of the matrix representation of the Hamiltonian except the diagonal element. While the structure returned by \inline{offdiagonals} can be used as a Julia array, it is lazily generated on the fly and as such does not require allocating any memory on the heap. The other way to access off-diagonal elements is through random sampling:
\begin{jlcodeblock}
random_offdiagonal(column)
# (
#     BoseFS(1, 1, 1, 1, 1, 1, 1, 1, 0, 2, 1, 1),
#     0.041666666666666664,
#     -1.4142135623730951,
# )
\end{jlcodeblock}
This returns the connected address, the probability with which it was selected, and its value.
The functions presented here give us enough information about the Hamiltonian to perform exact and stochastically sampled matrix-vector products.

\subsubsection{Modified Hamiltonians}\label{ssec:ham-wrappers}

A major benefit of using Julia for our implementation language is its multiple dispatch paradigm, which allows us to define efficient wrappers over existing data structures to modify their behaviour at essentially zero additional computational cost. One of the ways we make use of this in Rimu.jl is for transformations of Hamiltonians. This includes importance sampling (a similarity transformation) and symmetry transformations as discussed above in Sec.~\ref{sec:pcb}, or the results of simple mathematical operations on operators. Since many useful transformations have a similar structure and can be implemented as simple changes to the nonzero matrix elements of an operator, Rimu.jl provides a mechanism for defining them via the \inline{ModifiedHamiltonian} interface.

The full interface is described in the online documentation and more elaborate examples can be found by reading the Rimu.jl source code, e.g.~of \inline{ParitySymmetry}.
Here we demonstrate the implementation of a simple example of a scaled Hamiltonian that implements the operation of scaling a real-valued Hamiltonian with a real constant $\alpha$, i.e.~$\tilde{H} = \alpha \hat{H}$. This can be implemented with Rimu.jl as follows:\footnote{A slightly more elaborate version of \inline{ScaledHamiltonian} is already implemented in Rimu.jl. Here we show a simplified implementation focussing on the essentials.}
\begin{jlcodeblock}
import Rimu.Hamiltonians: ModifiedHamiltonian, modify_diagonal,
    modify_offdiagonal, parent_operator, LOStructure, AbstractHamiltonian

"""
    ScaledHamiltonian(H::AbstractHamiltonian, α::AbstractFloat)

The product of the Hamiltonian `H` with the real-valued scalar `α`.
"""
struct ScaledHamiltonian{
    T<:Real, H<:AbstractHamiltonian{T}
} <: ModifiedHamiltonian{T}
    hamiltonian::H
    α::T
end
function modify_diagonal(s::ScaledHamiltonian, source, value)
    return s.α * value
end
function modify_offdiagonal(s::ScaledHamiltonian, source, destination, value)
    return destination => s.α * value
end
parent_operator(s::ScaledHamiltonian) = s.hamiltonian
LOStructure(s::ScaledHamiltonian) = LOStructure(s.hamiltonian)
\end{jlcodeblock}
The main functions \inline{modify_diagonal} and \inline{modify_offdiagonal} work by altering the matrix element \inline{value} of the original Hamiltonian. More complicated examples can also make use of the \inline{source} and \inline{destination} addresses that indicate where in the matrix the element is located. It is further possible to reroute spawns by changing the returned \inline{destination} address. Setting the trait \inline{LOStructure} conveys information about the structure of the Hamiltonian (e.g.\ Hermitian, diagonal, etc.) that is used throughout Rimu.jl.
With these definitions, a \inline{ScaledHamiltonian} can be used just like any other Hamiltonian implemented in Rimu.jl. For example:
\begin{jlcodeblock}
hamiltonian = HubbardReal1D(BoseFS(1,0,1))
scaled_hamiltonian = ScaledHamiltonian(hamiltonian, 2.0)

Matrix(hamiltonian)
# 6×6 Matrix{Float64}:
#   0.0      -1.0      -1.0      -1.41421  -1.41421   0.0
#  -1.0       0.0      -1.0       0.0      -1.41421  -1.41421
#  -1.0      -1.0       0.0      -1.41421   0.0      -1.41421
#  -1.41421   0.0      -1.41421   1.0       0.0       0.0
#  -1.41421  -1.41421   0.0       0.0       1.0       0.0
#   0.0      -1.41421  -1.41421   0.0       0.0       1.0

Matrix(scaled_hamiltonian)
# 6×6 Matrix{Float64}:
#   0.0      -2.0      -2.0      -2.82843  -2.82843   0.0
#  -2.0       0.0      -2.0       0.0      -2.82843  -2.82843
#  -2.0      -2.0       0.0      -2.82843   0.0      -2.82843
#  -2.82843   0.0      -2.82843   2.0       0.0       0.0
#  -2.82843  -2.82843   0.0       0.0       2.0       0.0
#   0.0      -2.82843  -2.82843   0.0       0.0       2.0
\end{jlcodeblock}

\subsection{Dict Vectors}\label{ssec:dvecs}

Since the Hamiltonians we are interested in operate on vectors that are not indexed by integers but rather by Fock state addresses, standard (sparse) vectors are not a good fit for Rimu.jl. Instead, we use a dictionary-like structure to represent them. The underlying implementation based on hash tables provides us with $\mathcal{O}(1)$ access time when we need to look up a vector element when a Fock address is provided, i.e.\ independent of the number of stored elements. This is particularly important for the FCIQMC algorithm where contributions of positive and negative value need to be added to achieve what is called ``walker annihilation'' in the literature \cite{boothLinearscalingParallelisableAlgorithms2014}.

The available implementations of these generalised vectors fall under the \inline{AbstractDVec} supertype and generally behave like Julia dictionaries, but will return zero when indexed by an address that is not stored. They can be used in linear algebra operations involving \inline{AbstractHamiltonian}s, which were described in more detail in Sec.~\ref{ssec:hamiltonians}. Additionally, they hold a \inline{StochasticStyle}, which tells Rimu.jl how to apply an operator stochastically. Currently, there are three concrete implementations of vectors available: \inline{DVec}, \inline{InitiatorDVec}, and \inline{PDVec}.

The most basic form of \inline{AbstractDVec} is the \inline{DVec}, which is a thin wrapper over a standard Julia dictionary and only supports single-threaded operations. As an example, consider a \inline{DVec}, which is constructed with one or more \inline{address => value} pairs:
\begin{jlcodeblock}
dvec1 = DVec(BoseFS(1,1,1,1) => 1.0, BoseFS(2,0,2,0) => -1.0)
# DVec{BoseFS{4, 4, BitString{7, 1, UInt8}},Float64} with 2 entries, …
#   fs"|2 0 2 0⟩" => -1.0
#   fs"|1 1 1 1⟩" => 1.0
\end{jlcodeblock}
As mentioned above, it can be indexed much like a dictionary, but will return zero if the requested key is not stored.
\begin{jlcodeblock}
dvec1[BoseFS(1, 1, 1, 1)] # 1.0
dvec1[BoseFS(4, 0, 0, 0)] # 0.0
\end{jlcodeblock}
The vector can be multiplied with an operator using the standard multiplication function or its in-place counterpart \inline{LinearAlgebra.mul!}:
\begin{jlcodeblock}
hamiltonian = HubbardMom1D(BoseFS(1,1,1,1))
# HubbardMom1D(fs"|1 1 1 1⟩"; u=1.0, t=1.0)

dvec2 = hamiltonian * dvec1
# DVec{BoseFS{4, 4, BitString{7, 1, UInt8}},Float64} with 16 entries, …
#   fs"|4 0 0 0⟩" => -0.612372
#   fs"|2 2 0 0⟩" => 1.0
#   fs"|3 0 1 0⟩" => 0.612372
#   fs"|1 2 1 0⟩" => -0.707107
#   fs"|2 0 2 0⟩" => -2.5
#   fs"|0 2 2 0⟩" => 1.0
#   fs"|1 0 3 0⟩" => 0.612372
#   fs"|0 0 4 0⟩" => -0.612372
#   fs"|2 1 0 1⟩" => -0.353553
#   fs"|0 3 0 1⟩" => 0.612372
#   fs"|1 1 1 1⟩" => 3.0
#   fs"|0 1 2 1⟩" => -0.353553
#   fs"|2 0 0 2⟩" => 1.0
#   fs"|1 0 1 2⟩" => -0.707107
#   fs"|0 0 2 2⟩" => 1.0
#   fs"|0 1 0 3⟩" => 0.612372

dvec3 = similar(dvec1)
# DVec{BoseFS{4, 4, BitString{7, 1, UInt8}},Float64} with 0 entries, …

mul!(dvec3, hamiltonian, dvec1)
dvec3 == dvec2 # true
\end{jlcodeblock}
Hamiltonians together with dict vectors (see Sec.~\ref{ssec:hamiltonians}) allow for a wide range of linear algebra operations. The list includes (but is not limited to) vector-operator-vector dot-products, matrix-vector multiplication (and its in-place version with \inline{mul!}).
\begin{jlcodeblock}
dot(dvec1, dvec2) # 3.0
dot(dvec1, hamiltonian, dvec1) # 3.0
norm(dvec2) # 3.8078865529319543
sum(dvec2) # 9.449489742783177
\end{jlcodeblock}

\subsection{Parallel operations on vectors}\label{ssec:parallel}

The remaining implementation of \inline{AbstractDVec} is \inline{PDVec}, which allows for parallel and distributed operations and is the default choice on multi-core and/or distributed computers.

\begin{figure}[ht!]
\centering
\includegraphics[width=.8\textwidth]{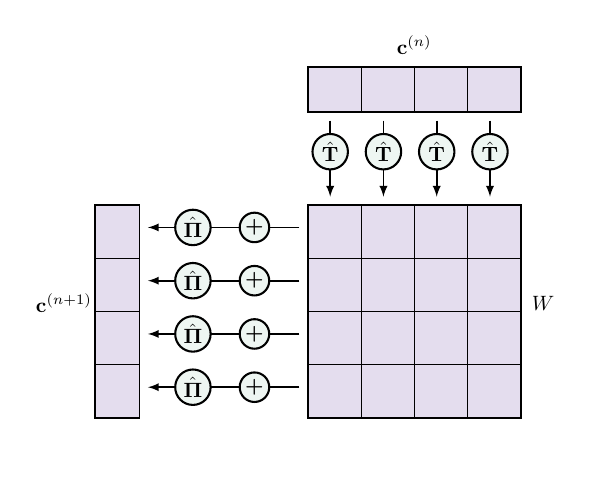}
\caption{\label{fig:pdvec}A diagram showing a parallel (multi-threaded) operator-vector product $\check{\vb{T}}\vb{c}^{(n)}$ followed by stochastic vector compression $\hat{\vb{\Pi}}$ as relevant for the FCIQMC algorithm of Eqs.~\eqref{eq:fciqmc-stochastic} and \eqref{eq:stochasticTransitionOperator}. Segmentation of the working memory allows for fully parallel processing without race conditions as each core operates on its own memory segment. The operation is performed in two phases. First, the operator $\check{\vb{T}}$ is applied to each segment of $\vb{c}^{(n)}$ with the result stored column-wise in the working memory $W$. Then, the segments in $W$ are summed row-wise, (optionally) compressed, and stored to the result vector $\vb{c}^{(n+1)}$. The diagram shows a vector split into four segments, which would allow both phases to be performed in parallel on four CPU threads.}
\end{figure}

\begin{figure}[ht!]
\centering
\includegraphics[width=.5\textwidth]{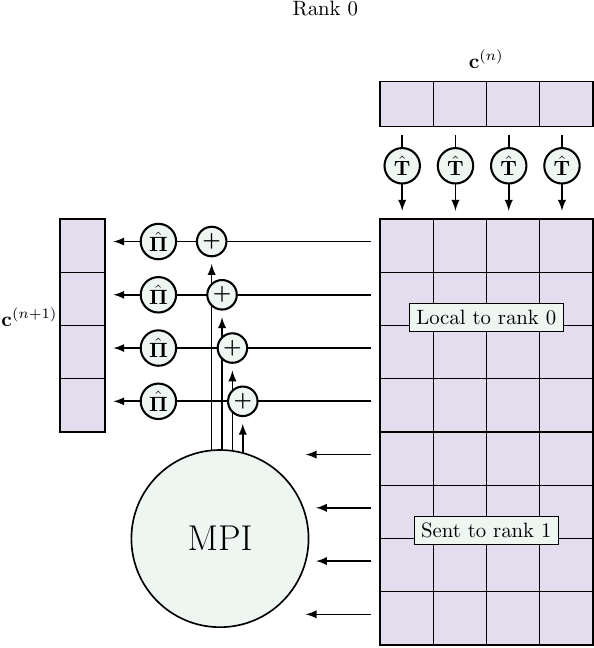}\includegraphics[width=.5\textwidth]{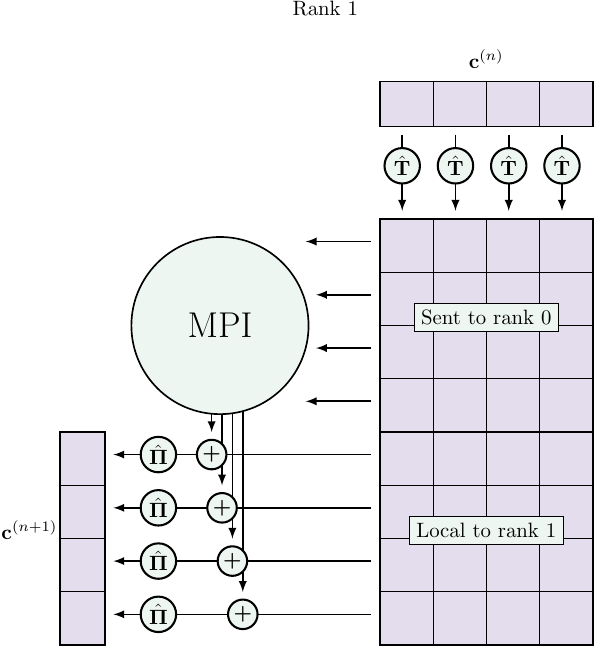}
\caption{\label{fig:pdvec-mpi}A diagram showing a distributed (stochastic) operator-vector product $\check{\vb{T}}\vb{c}^{(n)}$. The vector is split into chunks that are distributed over several (here two) ranks (nodes). Then after the operator $\check{\vb{T}}$ has been applied to each segment of $\vb{c}^{(n)}$, the results are stored column-wise in the working memory. The segments are then distributed among all MPI ranks, summed row-wise and (optionally) compressed, and stored to the result vector $\vb{c}^{(n+1)}$. The diagram shows the process on an example that uses two ranks with four threads each.}
\end{figure}

For most use-cases, a \inline{PDVec} is used in the same manner as a \inline{DVec}, while many operations performed on it are automatically parallelised. Parallelisation is achieved by splitting the storage of the vector into $k$ segments (each being a Julia dictionary) where $k$ is the number of threads available on the system. Then, each (address, value) pair is stored in the segment determined by the hash function of the address. Mathematically, we can think of a \inline{PDVec} $\vb{p}$ as being split into subvectors $\vb{p}^{(i)}$ with disjoint support
\begin{equation}
    \vb{p} = \sum_{i=1}^k \vb{p}^{(i)}\,.
\end{equation}
Then, any element-wise operation on $\vb{p}$ including reductions can be performed on each subvector independently. This allows for parallelisation of many functions, such as sums, dot products, vector compression, or any operation defined in terms of \inline{mapreduce}. The matrix-vector multiplication, can also be performed segment-wise as
\begin{equation}\label{eq:segment-matmul}
    \vb{T}\vb{p} = \sum_{i=1}^k\vb{T}\vb{p}^{(i)}\,.
\end{equation}
However, since the supports of $\vb{T}\vb{p}^{(i)}$ are no longer disjoint, performing the sum is not possible until all threads are finished. To avoid this issue, a working memory data structure $W$ is introduced. $W$ is a $k\times k$ array of dictionaries in which each of its rows and columns is organised in the same way as the vector $\vb{p}$. The intermediate result $\vb{T}\vb{p}^{(i)}$ is stored in the $i$-th column of the working memory. Finally, to perform the sum in Eq.~\eqref{eq:segment-matmul}, the rows of $W$ are summed, which can again be done in parallel. A schematic of this process is shown in Fig.~\ref{fig:pdvec} including the stochastic operations required for the FCIQMC algorithm. Deterministic matrix-vector multiplication proceeds in an analogous way omitting the stochastic compression.

Rimu.jl also supports distributed computing using the MPI~\cite{mpi41} standard through \href{https://github.com/JuliaParallel/MPI.jl}{MPI.jl}~\cite{byrneMPIjlJuliaBindings2021}. In a distributed scenario where there are $r$ MPI ranks with $k$ threads each, the storage of a \inline{PDVec} is divided into $rk$ segments, of which $k$ are stored on each rank. Performing operations on the vector is similar to as described above, except that MPI synchronisation is performed on every reduction or matrix-vector multiplication, and the working memory is now an $rk\times k$ array. A schematic is shown in Fig.~\ref{fig:pdvec-mpi}.

\section{Numerical results}\label{sec:numerical-results}

\subsection{Semistochasticity and vector compression styles}\label{ssec:semistochastic}

In this section, we discuss how changing the details of the matrix and vector compression algorithms affects the result of an FCIQMC calculation. Specifically, we look at two ways the algorithm described in Sec.~\ref{ssec:fciqmc:montecarlo} can be modified in Rimu.jl.

To provide a quantitative comparison, we will look at a measure of statistical efficiency, which is based on one introduced in Ref.~\cite{greeneWalkersStochasticQuantum2019a}. We define it as
\begin{equation}
  f = {\left(\sigma_{\bar{S}}^2\Omega \tilde{t}\right)}^{-1}
\end{equation}
where $\sigma_{\bar{S}}$ is the (blocked) error estimate of the mean of the shift, $\Omega$ is the number of time steps (not including the equilibration phase), and $\tilde{t}$ is the median of the wall time taken to perform one FCIQMC step. Note that this definition differs from the one defined in Ref.~\cite{greeneWalkersStochasticQuantum2019a} in the inclusion of the median of the wall-time in the measure. We include it because changing the compression algorithm often results in better statistics at a cost of more CPU time required per step. All measurements in this section were performed as single-threaded computations.

\emph{Late and early compression ---} In Sec.~\ref{ssec:fciqmc:montecarlo}, we
apply the stochastic vector projection operator $\check{\vb{\Pi}}$ after the stochastic operator $\check{\vb{T}}$ has been sampled and the contributions from the different parts of the input vector $\vb{c}$ have been summed, \ie
\begin{equation}\label{eq:late-comp}
  \check{\vb{T}}\vb{c} = \check{\vb{\Pi}}\left(\sum_i\tilde{\vb{T}}_{:,i}c_i\right)\,.
\end{equation}
We will refer to this approach as late compression. An alternative (standard~\cite{boothFermionMonteCarlo2009,gutherNECINElectronConfiguration2020}) approach, which we call early compression, is to apply the compression operator to each spawn individually as
\begin{equation}\label{eq:early-comp}
  \check{\vb{T}'}\vb{c} = \sum_i\check{\vb{\Pi}}\tilde{\vb{T}}_{:,i}c_i\,.
\end{equation}
These approaches are equivalent on average. The only difference between them is whether annihilation is done before or after compression. Because of this, early compression is more efficient to compute, while late compression produces better statistics.

\begin{figure}[t!]
\includegraphics[width=\textwidth]{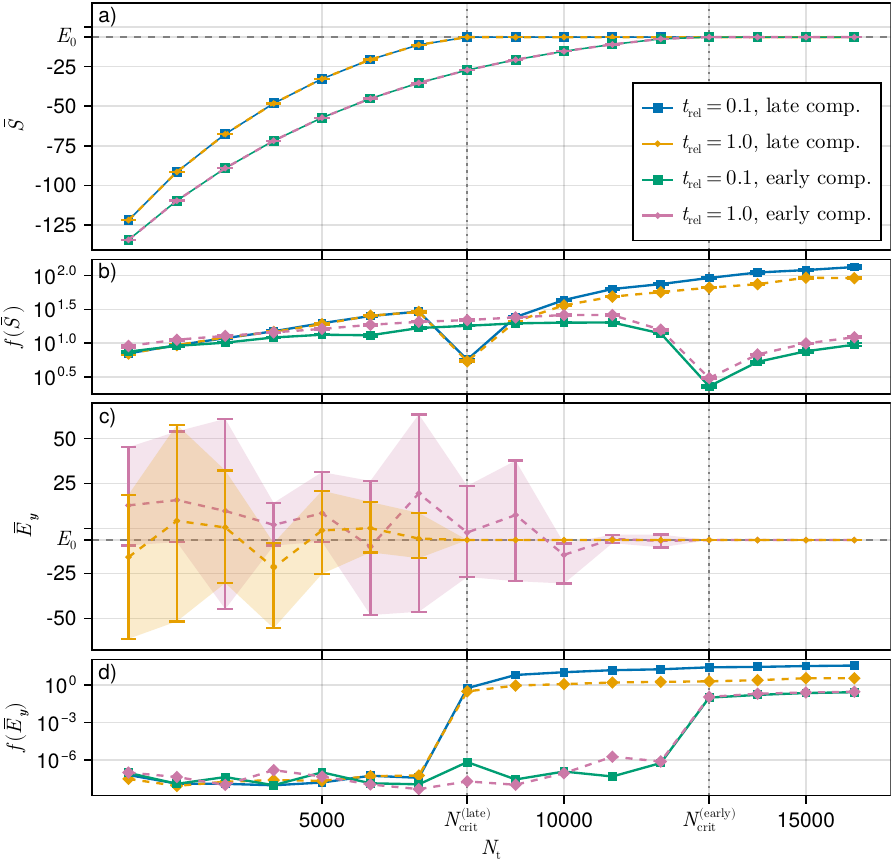}
\caption{\label{fig:signproblem}The effect of late compression on the sign problem. The plot shows (a) the mean of the shift $\langle S\rangle$  and (b) the statistical efficiency of the shift $f(\langle S\rangle)$. Panels (c) and (d) show the projected energy and corresponding statistical efficiency, respectively. The FCIQMC computations were performed with a momentum space Bose Hubbard model with $N=10$ particles in $M=10$ momentum modes with strong interactions at $u=6$ and a hopping strength of $t=1$. Each point in the plots is an average obtained from a time series of $10^7$ steps at $\dt= 10^{-3}$, where the first $10^5$ steps were discarded.
$t_\mathrm{rel}$ is the semistochastic threshold described in the text (see also Fig.~\ref{fig:efficiency}).
}
\end{figure}

\begin{figure}[t!]
\includegraphics[width=\textwidth]{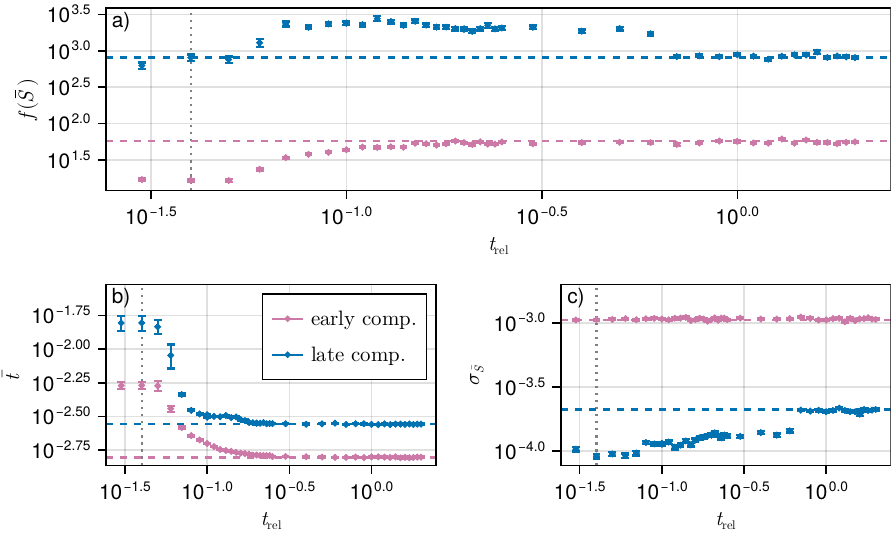}
\caption{\label{fig:efficiency}The effect of the relative semistochastic threshold $t_{\mathrm{rel}}$ and late compression on (a) the statistical efficiency of the shift $f(\langle S\rangle)$, (b) the median time per step $\bar{t}$, and (c) the (blocked) error bar size $\sigma_{\bar{S}}$. The FCIQMC computations were performed with a real space Bose Hubbard model with $N=12$ particles in $M=12$ modes with strong interactions at $u=6$ and a hopping strength of $t=1$. The horizontal dashed lines represent a setting of $t_{\mathrm{rel}}=\infty$, \ie no semistochasticity. The gray vertical dotted line represents a threshold where all of the spawns in the FCIQMC run are performed exactly.
}
\end{figure}

The results of comparisons between the two vector compression algorithms are shown in Figs.~\ref{fig:signproblem} and~\ref{fig:efficiency}. We note that in all cases, applying compression late has the effect of reducing error bar size, while increasing the runtime per step. These competing effects appear to provide an overall benefit when the FCIQMC instantaneous vector is peaked significantly. This can be explained by the fact that there, configurations in the vector with small coefficient values are allowed to spawn to larger ones uncompressed. On top of that, several small spawns hitting the same configuration are compressed less severely than they would have been otherwise.

When the system has a sign problem (Fig.~\ref{fig:signproblem}), the difference between the two methods is dramatic. There, we notice that by applying late compression, the sign problem is overcome at a smaller value of $N_\mathrm{t}$ and the statistical efficiency is increased considerably. Reducing the relative semistochastic threshold $t_{\mathrm{rel}}$ increases statistical efficiency further, but does not appear to affect the sign problem at all. Panels (c) and (d) show a similar result with projected energy $\bar{E}_{\vb{y}}$ with a simple single-configuration projector. There, we see the error bars drop drastically once the sign problem is overcome, while the statistical efficiency follows a pattern similar to what is shown in panel b). In such cases, applying compression late is strictly better than applying it early.

\emph{Semistochastic threshold ---} In Eq.~\eqref{eq:column-compression}, we note that we perform an exact spawning step if the coefficient value $c_i$ is larger than the number of non-zero off-diagonal elements in a column of the matrix $\vb{T}$. This can be modified by introducing a semistochastic threshold $t_{\mathrm{rel}}$ and performing the exact step if $c_i t_{\mathrm{rel}} \ge m_i$. Setting a lower value $t_{\mathrm{rel}}$ forces the algorithm to take more exact steps, which reduces the noise in the algorithm, while setting a higher value is more time-efficient. These competing effects cancel favourably, and a larger statistical efficiency is achieved, as shown in Fig.~\ref{fig:efficiency}. However, this is highly Hamiltonian dependent and the effects are not always as dramatic as those presented in Fig.~\ref{fig:efficiency}. The default value of $t_{\mathrm{rel}}=1$ often works well enough in practical situations.

\subsection{Performance benchmarks}

In this section, we present performance benchmarks that demonstrate how FCIQMC computations scale when a large number of threads or computer nodes is utilised. All presented benchmarks were performed with the two-dimensional Bose--Hubbard model with $N=36$ particles in $M=6\times 6$ sites, an interaction strength of $u=1$ and a hopping strength of $t=0.1$ modified with a Gutzwiller importance sampling ansatz~\cite{ghanemPopulationControlBias2021}. This is a system in the superfluid regime, where the ground state wavefunction is spread out across the Hilbert space, making it a somewhat difficult problem for FCIQMC to solve. The dimension of the Hamiltonian in question is $7.97\times 10^{21}$.

\begin{figure}[t!]
\includegraphics[width=\textwidth]{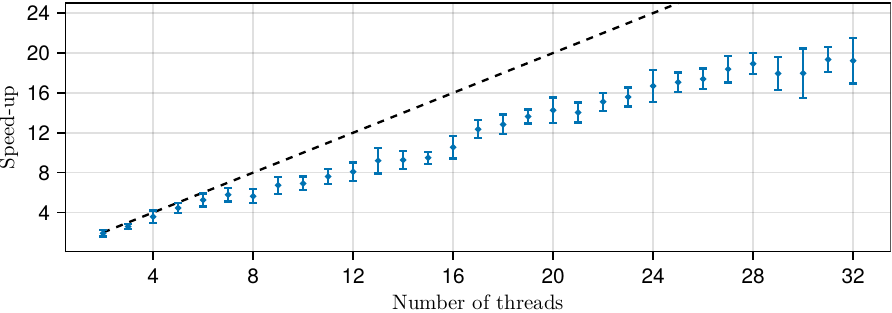}
\caption{\label{fig:multithreading}Scaling of shared memory parallelism. The speed-up relative to one thread for a multithreaded FCIQMC computation. The dashed black line shows the ideal linear scaling. In this regime, the speed-up is roughly linear with a slope of 0.61.}
\end{figure}

\emph{Single-node multithreading ---} Figure \ref{fig:multithreading} shows the result of the first benchmark, where we measure the multithreading scaling on a single computer node. The number of walkers was set to $N_{\mathrm{t}}=10^{7}$. In each benchmark, we perform a 25{,}000-step equilibration, after which we measure the wall-time it takes to perform $50{,}000$ steps. Then, each such benchmark is repeated 8 times, and the average and standard deviation between the 8 runs is reported.
As seen in Fig.~\ref{fig:multithreading}, the scaling is favourable -- 32 threads presents a 20-fold speed-up compared to a single-threaded computation, performing over 112 steps per second on average.

\emph{Multi-node parallelism ---} For measuring how Rimu.jl performs in a multi-node distributed setting, we employed the resources of the Frontier supercomputer at the Oak Ridge National Laboratory. There, each compute node consists of a 64-core AMD Optimized 3rd Gen EPYC CPU, with 56 allocatable cores under the so-called low-noise mode layout, \ie 8 CPUs are reserved for system processes.
We have performed the benchmarks on the same $6\times 6$ Bose-Hubbard model described above, but have used much larger numbers of walkers, up to $5.6\times 10^9$ in total.


\begin{figure}[t!]
    \centering
    \includegraphics[width=0.45\linewidth]{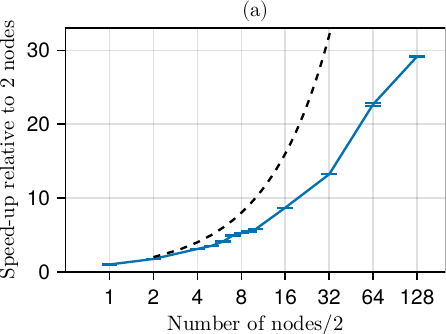}
    \includegraphics[width=0.45\linewidth]{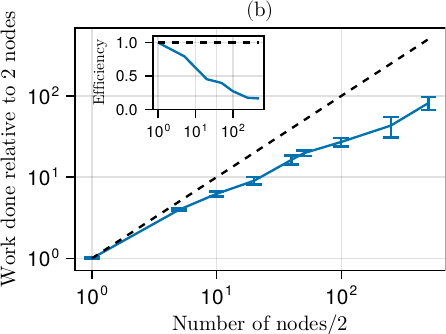}
    \caption{Scaling of multi-node parallelism. (a) The speed-up relative to 2 nodes of the average wall-time for performing a single FCIQMC step with Rimu.jl with $10^9$ walkers using up to 256 nodes with $14{,}336$ processors. The dashed line shows the ideal linear scaling.
    (b) The weak-scaling tests from the average wall-time for performing a single FCIQMC step. The main plot shows the relative work done, which is the number of walkers processed per time unit compared to the computation running on two nodes. The dashed line shows the ideal scaling.
    The inset shows the scaling efficiency
    calculated as (walkers processed on $N$ nodes)/($N/2\, \times$ walkers processed on 2 nodes) in a fixed amount of time. 
    }
    \label{fig:scaling}
\end{figure}

\emph{Strong scaling ---} For the first test, we look at strong scaling. It measures how using more computing power affects the time it takes to get a solution. Ideally, increasing the computational resources by a factor of $N$ would yield an $N$-fold speed-up. We present the results of benchmarks performed with $10^9$ walkers in Fig.~\ref{fig:scaling}(a). While the obtained speed-up trails the ideal scaling, Rimu.jl's performance maintains a general upward trajectory with no visible plateau.

\emph{Weak scaling ---} In the second test, we look at weak scaling. Here, both the size of the problem and the computational resources increase at the same rate. An algorithm that scales ideally would take the same amount of time for all samples taken this way. In this case, we have set Rimu.jl to use $5{,}600{,}000$ walkers per node and we have run the benchmarks on up to 1000 nodes. In Fig.~\ref{fig:scaling}(b), we present the scaling results.
We again see less than ideal scaling, however the amount of work done is increasing steadily even up to 1000 nodes. There, we see a six-fold slow-down, which corresponds to the efficiency dropping to around 0.16 for the largest calculations.

The observed reductions in efficiency for these very large calculations could be due to a number of different issues. Time spent on communication and copying data between the nodes is a potential contributing issue that will get worse when the number of nodes is increased. However, we assume that the most important issue is the unequal distribution of workload between the parallel nodes and processes. In order to perform walker annihilations in the FCIQMC algorithm, all nodes are synchronised in each time step, i.e.~all nodes must wait for the slowest. In the current implementation, the distribution of configuration-coefficient pairs to the nodes (determining the workload) for processing is done passively through a hashing process, i.e.~by (pseudo) random assignment.

The scaling benchmarks of this section show that significant performance gains could be obtained by optimising the code for large-scale computations. We expect that the largest improvement could be gained by an implementation of active load-balancing, which has improved the  efficiency for large-scale computations with similar codebases\cite{gutherNECINElectronConfiguration2020,spencerHANDEQMCProjectOpenSource2019}. However, for many research-grade projects, such large calculations are not necessary and single or few-node calculations suffice. The Rimu.jl codebase runs efficiently with such smaller computations running on a single or a small number of nodes. The results published in Refs.~\cite{taylorBoundExcitedStates2025,alhyderLatticeBosePolarons2025,backertEffectiveTheoryStrongly2025,rammelmullerMagneticImpurityOnedimensional2023,yangPolaronDepletonTransitionYrast2022} were all generated using Rimu.jl  with such smaller computations.

\section{Conclusion and outlook}\label{sec:conclusion}

We have presented Rimu.jl, a Julia package that implements quantum mechanical Hamiltonians and allows us to diagonalise them exactly or with the full configuration interaction quantum Monte Carlo (FCIQMC) algorithm. By utilising Julia's rich type system to encode the Fock states, Hamiltonians, and vectors, we are able to solve problems much larger than would have been possible with standard linear algebra algorithms. In addition, the code can be used on parallel and distributed systems. We have discussed the FCIQMC algorithm, key implementation details of the package, and presented benchmarks performed on a large-scale distributed system.

Rimu.jl currently supports Hamiltonians for cubic lattice problems in real and momentum space with on-site and extended interactions, continuum momentum space and transcorrelated~\cite{jeszenszkiEliminatingWavefunctionSingularity2020a} Hamiltonians, harmonic oscillator basis and the Fr\"ohlich polaron Hamiltonians. A \inline{MolecularHamiltonian} type for performing quantum chemistry calculations and a link to the quantum chemistry package \href{https://github.com/fkfest/ElemCo.jl}{ElemCo.jl} were recently added through a Google Summer of Code project~\cite{liAbInitioQuantumChemistry}. Some implemented observables include density-density correlations, reduced density matrices and momentum density. Future work could include new Hamiltonians such as spin-orbit coupled systems or new lattice geometries. Where these would require the definitions of new Hamiltonians or Fock state types, such additions are easy to make either as part of or outside of Rimu.jl due to the design of the package combined with Julia's multiple-dispatch and just-in-time compilation capabilities.
Implemented features of Rimu.jl include excited state calculation~\cite{bluntExcitedstateApproachFull2015a}, the initiator approximation~\cite{clelandCommunicationsSurvivalFittest2010}, importance sampling~\cite{ghanemPopulationControlBias2021,umrigarDiffusionMonteCarlo1993}, and thread-parallel and MPI-distributed~\cite{mpi41} calculations.
Future work may include extending the algorithm with modern features like the adaptive shift method~\cite{ghanemAdaptiveShiftMethod2020}, preconditioning~\cite{bluntPreconditioningPerturbativeEstimators2019a,neufeldAcceleratingConvergenceFock2020a}, or improved load balancing. A Rimu.jl extension for importance sampling and variational Monte Carlo is available through the (currently unregistered) \href{https://github.com/RimuQMC/Gutzwiller.jl}{Gutzwiller.jl} package.

The package is under active development with new Hamiltonians, observables, and features being added regularly. Please consult the \href{https://RimuQMC.github.io/Rimu.jl/stable/}{project documentation} for an up-to-date list of available Hamiltonians and features. The project is open-source and is licenced under the MIT licence. Contributions from the community in the form of issues or pull requests with code for bug fixes or new features are welcome.

\section*{Acknowledgements}

The authors would like to thank all contributers to the Rimu.jl GitHub repository, in particular Satyanand Kuwar, Jamie Taylor, Yosef Rabinowitz, Chuhao Li, Rohan Kumar, Alex Pletzer, Emma Wang, Tilak Patel, and Chris Scott for their contributions to the code and Ali Alavi and Daniel Kats for inspiration and helpful discussions.


\paragraph{Funding information}
M.Č., C.J.B., and J.B.\ received funding from the Marsden Fund of New Zealand (Contract No.\ MAU2007 and MPF-MAU2505), from government funding managed by the Royal Society of New Zealand Te Apārangi. We acknowledge support by the New Zealand eScience Infrastructure (NeSI/REANNZ) high-performance computing facilities in the form of a merit project allocation and a software consultancy project.
This research used resources of the Oak Ridge Leadership Computing Facility at the Oak Ridge National Laboratory, which is supported by the Office of Science of the U.S.~Department of Energy under Contract No.~DE-AC05-00OR22725.



\bibliography{rimu_codebase.bib}

\nolinenumbers

\end{document}